\documentclass[format=acmsmall, review=false, screen=true]{acmart}

\usepackage{url}
\usepackage{caption}
\usepackage{graphicx}
\usepackage{float} 
\usepackage{subfigure}
\usepackage{subcaption}
\usepackage{balance}

\usepackage{pgfplots}
\pgfplotsset{compat=1.18}
\usepackage{enumitem}

\usepackage{booktabs} 
\usepackage{graphicx} 
\usepackage{tcolorbox} 
\usepackage{tabularx}
\usepackage{array}

\usepackage{tikz}
\usepackage{geometry}
\usetikzlibrary{shapes.geometric, arrows, positioning}

\tikzstyle{box} = [
    rectangle,
    draw,
    thick,
    minimum width=6cm,
    minimum height=3cm,
    text width=6cm,
    align=left,
    inner sep=8pt
]

\tikzstyle{arrow} = [->, thick]

\acmJournal{TOSEM}
\usepackage{algorithm}
\usepackage{algorithmic}
\usepackage{wrapfig}
\usepackage{multirow}
\usepackage{makecell}
\usepackage{utfsym}
\usepackage{listings}
\usepackage{xcolor}
\usepackage{colortbl}
\definecolor{codegreen}{rgb}{0,0.6,0}
\definecolor{codegray}{rgb}{0.5,0.5,0.5}
\definecolor{codeorange}{rgb}{1,0.49,0}
\definecolor{backcolor}{rgb}{0.95,0.95,0.96}

\usepackage{tcolorbox}
\tcbuselibrary{skins, breakable}
\usepackage{colortbl}
\usepackage{framed} 
\lstdefinestyle{mystyle}{
 backgroundcolor=\color{backcolor},
 commentstyle=\color{codegray},
 keywordstyle=\color{codeorange},
 numberstyle=\tiny\color{codegray},
 stringstyle=\color{codegreen},
 basicstyle=\ttfamily\footnotesize,
 breakatwhitespace=false,
 breaklines=true,
 captionpos=b,
 keepspaces=true,
 numbers=left,
 numbersep=5pt,
 showspaces=false,
 showstringspaces=false,
 showtabs=false,
 tabsize=2,
 xleftmargin=10pt,
}

\begin{document}

\title{Rule Taxonomy and Evolution in AI IDEs: A Mining and Survey Study}
\author{Guangzong Cai}
\orcid{0009-0003-7346-2871}
\affiliation{%
  \institution{School of Computer Science, Wuhan University}
  \city{Wuhan}
  \country{China}
}
\email{guangzongcai@whu.edu.cn}

\author{Ruiyin Li}
\orcid{0000-0001-8536-4935}
\affiliation{%
  \institution{School of Computer Science, Wuhan University}
  \city{Wuhan}
  \country{China}
}
\email{ryli_cs@whu.edu.cn}

\author{Peng Liang}
\orcid{0000-0002-2056-5346}
\affiliation{
  \institution{School of Computer Science, Wuhan University}
  \city{Wuhan}
  \country{China}
}
\email{liangp@whu.edu.cn}

\author{Zengyang Li}
\orcid{0000-0002-7258-993X}
\affiliation{%
  \institution{School of Computer Science, Central China Normal University}
  \city{Wuhan}
  \country{China}
}
\email{zengyangli@ccnu.edu.cn}

\author{Mojtaba Shahin}
\orcid{0000-0002-9081-1354}
\affiliation{%
  \institution{School of Computing Technologies, RMIT University}
  \country{Australia}
}
\email{mojtaba.shahin@rmit.edu.au}

\renewcommand{\shortauthors}{Cai et al.}

\begin{abstract}
The adoption of AI-powered Integrated Development Environments (AI IDEs) has introduced ``Rules'' as a novel software artifact, allowing developers to persistently inject project-specific constraints and architectural guidelines into the context of Large Language Models (LLMs). Despite their role in aligning AI behavior with developer intent, the taxonomy, evolution, and practical impact of these rules remain largely unexplored. To bridge this gap, we conducted a mixed-methods empirical study on AI IDE rules. By mining 83 open-source projects and extracting 7,310 rules, we established a comprehensive taxonomy comprising 5 primary and 25 secondary categories. We then triangulated these artifacts with survey responses from 99 practitioners. Our analysis identified a contrast between developer priorities and actual configurations: while practitioners rate architectural constraints as highly important, rule files in repositories primarily consist of low-level workflow and code formatting constraints. Furthermore, our analysis of 1,540 rule evolution events revealed that rules are updated frequently. Repository data further indicate that rule evolution is primarily driven by constructive context expansions (29.17\%) and enrichments (26.59\%). In contrast, surveyed developers reported modifying rules primarily to correct AI errors (77.78\%), typically by adding new negative constraints rather than editing existing ones. Finally, an artifact compliance assessment of 160 rule evolution events revealed that updating rules significantly improves the adherence of software artifacts, with the average artifact compliance rate increasing by 22.99\% (from 49.14\% to 72.13\%) following an update. Our study provides empirical insights that can help developers optimize prompting strategies and guide tool builders in designing automated conflict-detection and context-management mechanisms for AI IDEs.
\end{abstract}

\ccsdesc[500]{Software and its engineering~Integrated and visual development environments}
\ccsdesc[500]{Software and its engineering~Software development techniques}
\ccsdesc[500]{Software and its engineering~Software configuration management and version control systems}

\keywords{AI IDEs, Rules, Prompt Engineering, Agentic Coding}

\maketitle

\section{Introduction}\label{sec_introduction}
In recent years, the rapid advancement of Large Language Models (LLMs) has reshaped the paradigms of software engineering practice \cite{Ta2024, PeKaCiDe2023, PaGrMaNaMaMeZhFeCh2025}. Evolving from early code completion tools to today's AI Integrated Development Environments (AI IDEs) - such as Cursor \cite{Cursor}, Windsurf \cite{Windsurf}, and Trae \cite{Trae} - AI has transcended its role as a mere plugin, emerging as a coding agent capable of perceiving project context and executing complex tasks \cite{LiZhHa2025}. At the core of this transformation is a deepening interaction between developers and AI: developers no longer settle for one-off Q\&A sessions, but instead expect AI to strictly adhere to specific project conventions, architectural patterns, and coding styles \cite{YaLiLvDeGuJiLiLiLuLu2025}. To achieve this, many AI IDEs have introduced a novel software artifact called \textit{Rules} (e.g., Cursor~\cite{CursorRules}, Windsurf~\cite{WindsurfRules}, Trae~\cite{TraeRules}, Qoder~\cite{QoderRules}, Kiro~\cite{KiroRules}). By configuring specific rule files (e.g., \texttt{.cursor/rules}) within a project, developers can persistently inject project-specific knowledge (e.g., ``\textit{Please use Spring Boot version 4.0 or higher}'', ``\textit{Follow Domain-Driven Design}'') into the context of AI IDEs \cite{NaMaHeVaMy2024, RoMaHoMuWe2023}. This mechanism essentially functions as persistent prompt engineering \cite{VaKiAcLeMhGlAr2025, ShDa2026}, ensuring that the AI IDEs' behavior not only maintains long-term alignment with developer intent \cite{ChLiKaReThLeRuMaAdHa2025} but also explicitly enhances the quality of the generated code \cite{DeLaPa2025, KhDeMoLe2025, PiFaBiStEn2025}. Such a practice is gaining wide attention across the software development landscape \cite{LiZhHa2025, ChLiKaReThLeRuMaAdHa2025, ShDa2026}. In projects developed with AI IDEs, rule files are emerging as a critical type of configuration file \cite{StackOverflow2025, 2RoMaDeHoZa2026}. 

Despite the growing adoption of AI IDE rules, the understanding of their taxonomic characteristics and evolutionary patterns remains limited. First, as an emerging type of software artifact, the taxonomic characteristics of rules remain ill-defined. While a handful of studies have touched upon rule categorization, they are often limited by small sample sizes \cite{ChThReKaLeRuMaIi2025, ShDa2026} or restricted to specific IDEs \cite{SaCoMoVa2025, ChLiKaReThLeRuMaAdHa2025}, lacking a comprehensive and generalizable taxonomy. Second, rules are not static artifacts. As software projects iterate, codebases expand, and tech stacks update - compounded by the potential for AI ``hallucinations'' - developers must continuously maintain and refine these rules \cite{SeMaChSe2026}. However, the content of rule evolution, the reasons for its evolution, and how the compliance rate of software artifacts changes after rule evolutions remain largely unexplored. Without a deep understanding of their evolutionary lifecycle, rule files may become obsolete due to delayed updates or, worse, provide misleading context to AI IDEs, thereby undermining the effectiveness of human--AI collaboration in AI-assisted software development.

To bridge the aforementioned gaps, this work aims to conduct a large-scale empirical study to investigate the state of practice regarding AI IDE rules in open source software (OSS) projects. Specifically, this study is dedicated to addressing the following two Research Questions (RQs):

\begin{itemize}
    \item \textbf{RQ1}: What are the categories of rules in OSS projects developed by AI IDEs?
    \item \textbf{RQ2}: How do rules in OSS projects developed by AI IDEs evolve?
\end{itemize}

To answer these two RQs, we employed a mixed-methods research design, combining mining software repositories with a subsequent developer survey. First, utilizing the GitHub API combined with manual filtering, we curated a dataset of 83 projects explicitly developed using AI IDEs. From these projects, we collected 325 rule files, extracting a total of 7,310 individual rules. Through Open Coding \cite{Se1999}, we constructed a hierarchical taxonomy consisting of 5 major categories and 25 sub-categories. Next, we extracted rules that had undergone changes in their commit history. By analyzing rule-content diffs, co-changed files, and commit messages, we characterized the types of rule evolution, the reasons behind these changes, and the resulting shifts in software artifact compliance (i.e., the extent to which artifacts adhere to the updated rules). Furthermore, we designed and conducted a survey targeting developers with practical experience in configuring AI IDE rules. The survey consisted of 11 questions covering respondents' demographics and rule usage practices. We finally received 99 valid responses. By combining ``objective mining data'' with ``subjective developer perceptions'', we integrated these complementary perspectives to obtain a comprehensive understanding of rule taxonomy and evolution in AI IDEs.

The main \textbf{contributions} of this work are as follows:
\begin{itemize}
    \item We established a taxonomy of AI IDE rules (5 primary, 25 secondary categories) based on 7,310 mined rules. Triangulating this with a 99-participant survey revealed a clear contrast between developers' perceived architectural priorities and their actual formatting-centric configurations.
    \item We characterized rule evolution by analyzing 1,540 change events, identifying a gap in the perceived reasons for rule evolution between commit data (driven by context expansion) and developer responses (focused on correcting AI errors).
    \item We quantified the impact of rule maintenance on codebase adherence, demonstrating through an assessment of 160 rules that rule updates improved average artifact compliance by 22.99\%.
\end{itemize}

\textbf{Paper organization}: Section~\ref{sec_background} provides a brief background on AI IDEs and their rules. Section~\ref{sec_relatedwork} reviews related work on AI IDEs and rules in AI IDEs. Section~\ref{sec_design} outlines the research questions and details our data collection procedures and research methodology. Section~\ref{sec_result} presents the study results. Section~\ref{sec_discussion} discusses the findings and provides implications for both practitioners and researchers. Section~\ref{sec_validity} addresses the threats to validity. Section~\ref{sec_conclusion} concludes the work with future directions.

\section{Background}\label{sec_background}
This section first introduces the emergence of AI IDEs (Section~\ref{BKAIIDEs}) and then discusses the concept of rules in AI IDEs (Section~\ref{BKRules}), which serves as the primary mechanism for customizing their behavior.

\subsection{AI IDEs}\label{BKAIIDEs}
Unlike traditional IDEs that integrate AI features as optional extensions (e.g., VS Code with the GitHub Copilot plugin \cite{Copilot}), an AI IDE refers to an integrated development environment built with artificial intelligence as a core architectural component \cite{AIIDE}. In these IDEs, AI is not an auxiliary feature layered on top of an existing editor; rather, it constitutes the central organizing principle that shapes the IDE's functionalities, interaction paradigms, and iterative evolution \cite{YaLiLvDeGuJiLiLiLuLu2025}. This architectural distinction leads to fundamental differences in the assistant capability of AI IDEs. Plugin-based solutions operate within the constraints imposed by the host IDE, limiting their access to certain runtime states, internal editor signals, or fine-grained interaction data. In contrast, AI IDEs integrate Large Language Models (LLMs) directly into the core editor infrastructure, enabling deeper access to project-wide context, internal state transitions, and user interaction history. As a result, AI IDEs typically support advanced capabilities beyond code completion, including autonomous task decomposition, multi-file refactoring, architectural reasoning, context-aware code generation, and continuous alignment with project-specific configurations (e.g., rule files) \cite{TeVeRoBl2025}.

In this study, we focus on five representative AI IDEs that support user-defined rule configurations: Cursor \cite{Cursor}, Windsurf \cite{Windsurf}, Trae \cite{Trae}, Qoder \cite{Qoder}, and Kiro \cite{Kiro}. Table~\ref{tab:ai_ides} presents information on these AI IDEs, highlighting the locations of their rule files. These AI IDEs share a common characteristic: they allow developers to explicitly define ``Rules'' that the AI IDEs must follow during code generation and chat interactions. Notably, Kiro refers to its rule mechanism as Steering. By default, Kiro automatically generates three configuration files - \texttt{product.md}, \texttt{tech.md}, and \texttt{structure.md} - under the \texttt{.kiro/steering/} directory. Users may also define additional rule files within this directory to customize Kiro' behavior \cite{KiroRules}.

\begin{table}[t]
\centering
\caption{Comparison of AI IDEs and Their Rule Mechanisms}
\label{tab:ai_ides}
\small
\setlength{\tabcolsep}{4pt}

\begin{tabularx}{\linewidth}{p{2.1cm} p{1.7cm} p{2.3cm} >{\footnotesize\arraybackslash}X}
\toprule
\textbf{AI IDE} & \textbf{Provider} & \textbf{Release} & \textbf{Rule File Location(s)} \\
\midrule

\textbf{\includegraphics[height=0.32cm]{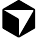} Cursor}
& Anysphere
& 6 Apr 2023 \cite{CursorChangelog}
& \texttt{.cursor/\allowbreak rules/}, \texttt{.cursorrules} \cite{CursorRules} \\ 

\addlinespace[3pt]

\textbf{\includegraphics[height=0.32cm]{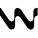} Windsurf}
& Codeium
& 13 Nov 2024 \cite{WindsurfChangelog}
& \texttt{.windsurf/\allowbreak rules/}, \texttt{.windsurfrules} \cite{WindsurfRules} \\

\addlinespace[3pt]

\textbf{\includegraphics[height=0.32cm]{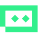} Trae}
& ByteDance
& 20 Jan 2025 \cite{TraeChangelog}
& \texttt{.trae/\allowbreak rules/} \cite{TraeRules} \\

\addlinespace[3pt]

\textbf{\includegraphics[height=0.32cm]{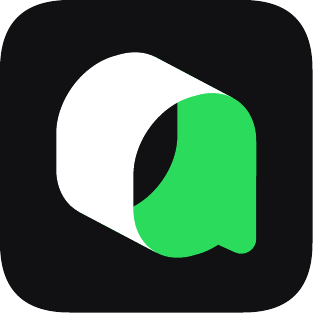} Qoder}
& Alibaba
& 21 Aug 2025 \cite{QoderChangelog}
& \texttt{.qoder/\allowbreak rules/} \cite{QoderRules} \\

\addlinespace[3pt]

\textbf{\includegraphics[height=0.32cm]{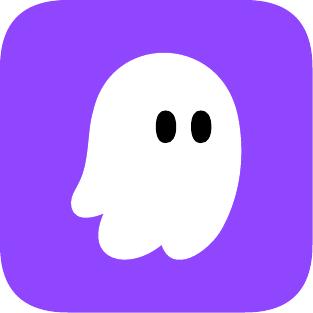} Kiro}
& Amazon
& 14 Jul 2025 \cite{KiroChangelog}
& \texttt{.kiro/\allowbreak steering/} \cite{KiroRules} \\

\bottomrule
\end{tabularx}

\vspace{3pt}
\footnotesize
\raggedright
\textit{Note:} The release dates are obtained from official changelogs. For Cursor, the listed date corresponds to the version in which it transitioned to the VSCode-based architecture. For Cursor and Windsurf, legacy rule file formats (e.g., \texttt{.cursorrules}, \texttt{.windsurfrules}) were later superseded by directory-based rule management mechanisms.

\end{table}


\subsection{Rules in AI IDEs}\label{BKRules}
In the context of AI IDEs, rules are textual instructions or directives defined by developers to constrain and guide the behavior of the LLM-driven AI assistant for software development~\cite{CursorRules, TraeRules}. These rules are typically aggregated within specific rule files located in the project's root or configuration directory. This mechanism functions as a form of persistent prompt engineering. When a user interacts with the AI assistant, the IDE automatically retrieves the relevant content from these rule files and injects them into the LLM's context window.

The primary purpose of rules is to align the AI IDEs' output with the specific requirements of the project, which often include architectural constraints, coding styles, library preferences, and development workflow guidelines \cite{ChLiKaReThLeRuMaAdHa2025, ShDa2026}. A typical rule file is written in natural language (often Markdown). Figure~\ref{fig:rule_example} shows a snippet of a typical rule file for a React Native project, sourced from the community rule repository \texttt{cursor.directory}\footnote{\url{https://cursor.directory/rules}}. As shown in Figure~\ref{fig:rule_example}, the rule explicitly defines the Role (``\textit{expert in TypeScript...}''), Style Constraints (``\textit{avoid classes}''), and Technical Constraints (``\textit{Use Expo's AppLoading}''). Without this rule, a generic LLM might generate class-based components or use deprecated libraries, requiring the developer to perform cumbersome manual corrections.

\begin{figure}[ht]
\begin{tcolorbox}[colback=gray!5!white,colframe=gray!75!black,title=Example of AI IDE Rules]
\texttt{You are an expert in TypeScript, React Native, Expo, and Mobile UI development.}

\textbf{Code Style and Structure}
\begin{itemize}
\item Write concise, technical TypeScript code with accurate examples.
\item Use functional and declarative programming patterns; avoid classes.
\end{itemize}

\textbf{Performance Optimization}
\begin{itemize}
\item Minimize the use of \texttt{useState} and \texttt{useEffect}; prefer context and reducers for state management.
\item Use Expo's \texttt{AppLoading} and \texttt{SplashScreen} for optimized app startup experience.
\end{itemize}
...
\end{tcolorbox}
\caption{An example of a rule file defining coding standards.}
\Description{A formatted example of an AI IDE rule file displayed inside a shaded box. The content includes a role definition stating expertise in TypeScript, React Native, Expo, and mobile UI development, followed by sections titled ``Code Style and Structure'' and ``Performance Optimization.'' Each section contains bullet-point guidelines, such as writing concise TypeScript code, using functional programming patterns, minimizing the use of useState and useEffect, and leveraging Expo-specific features for performance optimization.}
\label{fig:rule_example}
\end{figure}

\section{Related Work}\label{sec_relatedwork}

\subsection{Evaluation and Impact of AI IDEs}
Current research has extensively investigated AI-assisted programming tools, with a particular focus on AI IDEs \cite{YaLiLvDeGuJiLiLiLuLu2025, TeVeRoBl2025}. These studies can be broadly categorized into four dimensions: capabilities and benchmarking, impact on development practices, security assessments, and future directions.

\textbf{Capabilities and Benchmarks.} Initial studies have surveyed the landscape of AI coding tools, providing comparative analyses of features in AI IDEs such as Windsurf and Cursor \cite{TeVeRoBl2025}. Beyond functional comparisons, researchers have constructed datasets from GitHub to evaluate these AI IDEs \cite{LiZhHa2025, ZiWuBoBeGu2025} and explored their advanced capabilities in specific tasks, such as code refactoring \cite{HoLiKaAdIiHa2025}, issue resolution \cite{KuBaGuSoMu2025}, end-to-end microservice generation \cite{PeYiZhYaWaJiChLiRaZh2026}, and design issues in large-scale project generation~\cite{KaLiLiTaFeLiSh2026}. Due to its strong reasoning capabilities, Cursor is increasingly utilized as a baseline for tasks like codebase question answering \cite{YaTaChHe2025}, doc-to-code synthesis \cite{LiLiGuReHu2025}, and microservice bug localization \cite{OsYuBoAk2025}. 

\textbf{Impact on Development and Education.} A significant body of work investigates how AI IDEs reshape the software development lifecycle. While some studies highlight improvements in coding efficiency \cite{AgHeVa2026, BeRuBaRe2025}, others, such as He \textit{et al}. \cite{HeMiAgKaVa2026}, caution that Cursor may trade long-term code quality for short-term speed. To comprehend developer requirements and interaction behaviors, researchers have conducted questionnaire surveys on developers' experiences using AI coding assistants and CodeLLMs~\cite{VuPaHoKrPaMe2026} as well as empirical analysis of actual in-IDE human-AI conversation data \cite{TaChFaXuDhShMcHuLi2026}. In the educational domain, the impact of AI IDEs like Cursor on teaching and learning methodologies has been widely examined \cite{ScRu2025}.

\textbf{Security and Risk Assessment.} As AI IDEs gain widespread adoption, concerns regarding the security risks introduced by AI-assisted code generation have grown significantly. Security evaluations of AI IDEs have covered a wide spectrum of risks, including the vulnerability of generated code \cite{NiWaYaJiTaDaGaLiGuSo2025, LuZhLuLiHuChMaZhXiCa2025}, and security issues arising from the invocation of external tools (e.g., shell commands, file systems, and APIs) by AI IDEs \cite{XiLuLiZhZhLiLiChWaSh2025}. More sophisticated threats targeting AI IDEs and their context have also been reported, including prompt injection attacks \cite{LiZhLyZhWaLo2025}, malicious payloads embedded in configuration files \cite{LiLiLiZhWaPaJiWa2025}, and API key cracking \cite{NaScBr2026}.

\textbf{Emerging Paradigms and Future Directions.} With the rise of ``Vibe Coding'', a wave of recent research has focused on this novel programming paradigm, discussing the pivotal role of AI IDEs in enabling natural language-driven development \cite{ChQiWaWaYa2025, HuReLeXiHe2025, ChJiChWeJaZiJo2025, LiHoLiZhCaAl2025, GeMeDuLiZhWaWaYaLiCa2025}. Furthermore, researchers have discussed current challenges - such as limitations in the context window capacity of LLMs and hallucinations in AI-generated code by AI IDEs \cite{WaGoZhXuWa2025, 1RoMaDeHoZa2026} - while envisioning future directions like multimodal programming and self-evolving agents \cite{YaLiLvDeGuJiLiLiLuLu2025}.

\subsection{Rule Configuration for AI Coding Agents}
Research on AI coding rules can be categorized into two streams: studies focusing on specific functional aspects and empirical studies characterizing the rule ecosystem itself.

\textbf{Specific Applications and Impacts of Rules.}
Several studies have investigated the impact of rules in specific domains. In terms of efficiency, Lulla \textit{et al}. analyzed the impact of \texttt{AGENTS.md} on runtime performance and token consumption \cite{LuMoGaZhBaTr2026}. More critically, Gloaguen \textit{et al}. investigated the effect of repository-level context files (\texttt{AGENTS.md}) on automated agents' task success rates, revealing that auto-generated contexts yield no improvement, while human-written ones offer only marginal gains \cite{GlMuMuRaVe2026}. Regarding interaction and format, McMillan \textit{et al}. examined how file formats (e.g., YAML vs. Markdown) affect agent behavior \cite{Mc2026}, and further showed that configuration file design choices (e.g., length, structure, and instruction placement) have a negligible impact on instruction compliance. Instead, their findings indicate that an agent's adherence to rules is primarily affected by the specific type of coding task and the length of the interaction session; specifically, the likelihood of an agent complying with rules progressively degrades as it continuously generates more code (e.g., generating subsequent functions) within a single session \cite{Mc2026Instruction}. Vaithilingam \textit{et al}. proposed SemanticCommit, a mixed-initiative interface to facilitate intent integration into rule files \cite{VaKiAcLeMhGlAr2025}. Furthermore, Pimenova \textit{et al}. discussed the status of rules in the emerging ``Vibe Coding'' paradigm \cite{PiFaBiStEn2025}, and Hora \textit{et al}. provided the best practices for using rules to guide AI coding agents in generating mock objects and handling dependencies during software testing \cite{HoRo2026}.

\textbf{Empirical Characterization of Rules.}
A growing body of work has conducted large-scale empirical studies on rule files (e.g., \texttt{AGENTS.md}, \texttt{CLAUDE.md}, \texttt{.cursorrules}) to understand their structure and content. Galster \textit{et al}. analyzed configuration practices in agentic AI coding tools and observed that among the eight identified configuration mechanisms, static context files (e.g., \texttt{CLAUDE.md} and \texttt{AGENTS.md}) are the most widely adopted by developers in GitHub repositories. They dominate the configuration landscape and often serve as the sole mechanism used, whereas advanced mechanisms (such as Skills and Subagents) remain less prevalent, appearing in fewer than 20\% of repositories and primarily in Cursor-based projects \cite{GaMoLuAbTrBa2026}. Studies on Claude Code configurations \cite{SaCoMoVa2025, ChThReKaLeRuMaIi2025} found that these files typically exhibit a shallow hierarchical structure, with content predominantly focused on operational commands (build/run), implementation details, and high-level architecture. Studies on \texttt{AGENTS.md} \cite{SeMaChSe2026, ChLiKaReThLeRuMaAdHa2025} analyzed thousands of repositories, revealing that these files are not static documentation but evolving artifacts maintained through frequent updates. However, Chatlatanagulchai \textit{et al}. noted a significant gap: while functional-related rules are prioritized, non-functional rules like security and performance are rarely specified \cite{ChLiKaReThLeRuMaAdHa2025}. Jiang \textit{et al}. qualitatively analyzed 401 Cursor repositories. They constructed a taxonomy comprising five high-level themes (e.g., Norms, Guidelines) and explored how rule usage varies across programming languages and project domains \cite{ShDa2026}.

\subsection{Conclusive Summary}
Existing studies on AI IDEs have primarily examined their capabilities, impact on development practices, and associated security risks \cite{HeMiAgKaVa2026, NiWaYaJiTaDaGaLiGuSo2025}. These works typically treat AI IDEs as holistic tools or focus on the code artifacts they generate \cite{OsYuBoAk2025, HoLiKaAdIiHa2025}. In contrast, we focus on the Rule mechanism within AI IDEs, which connects ``developer intent'' with ``AI behavior''. By exploring the categories and evolution of rules, we reveal the lifecycle characteristics of this emerging software artifact, providing empirical evidence for developers to utilize rules more effectively and for AI IDE designers to optimize rule mechanisms.

Meanwhile, research on rule configuration has begun to explore the structure and usage of rule files \cite{SaCoMoVa2025, ChThReKaLeRuMaIi2025} and to propose initial taxonomies \cite{ShDa2026}. However, these studies are often limited to a single AI IDE and therefore do not capture the diversity of rule usage across different AI IDEs?. Furthermore, significant gaps remain regarding the dynamics of these rules. First, although existing studies (e.g., \cite{SeMaChSe2026, ChLiKaReThLeRuMaAdHa2025}) observed that rules evolve, their analysis was primarily descriptive, lacking a systematic investigation into the drivers behind these evolutions.
Second, no prior work has empirically evaluated how the compliance of software artifacts changes over time following rule evolution. Our study advances this field by not only constructing a comprehensive taxonomy of rules in AI IDEs but, more importantly, by employing multi-perspective data analysis to reveal why rules change and how the adherence of software artifacts shifts after rules are added or modified.

\section{Research Methodology}\label{sec_design}
\begin{figure*}[t]
\centering
\includegraphics[
    width=\textwidth,
]{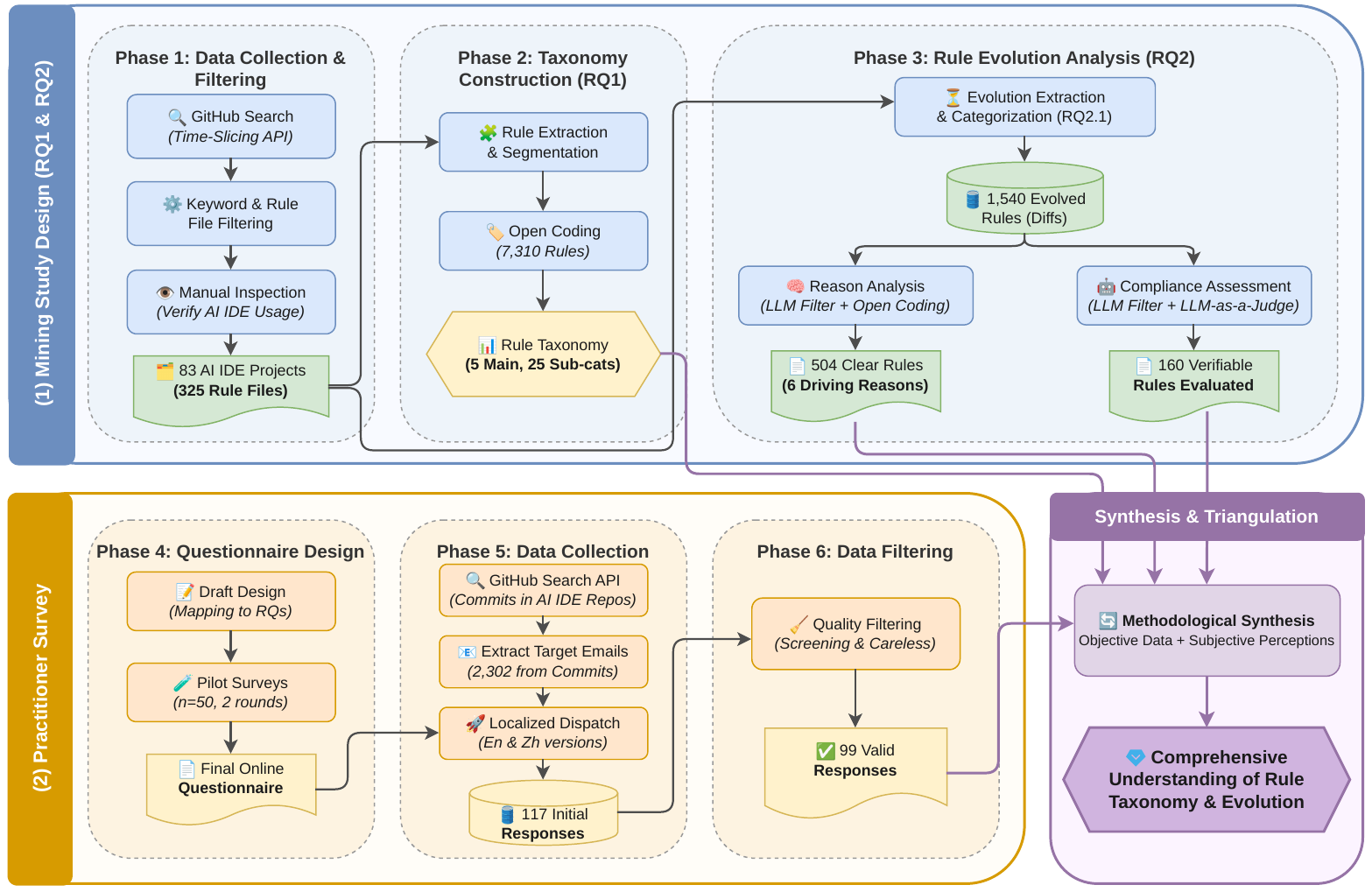}

\caption{Overview of the mixed-methods research process.}
\Description{This flowchart provides a comprehensive overview of the mixed-methods research process, divided into two main parallel tracks: ``(1) Mining Study Design'' at the top and ``(2) Practitioner Survey'' at the bottom. The process flows horizontally across six distinct phases, culminating in a final synthesis.  The Mining track spans three phases: Phase 1 involves data collection and filtering from GitHub, resulting in 83 AI IDE projects. Phase 2 covers taxonomy construction, segmenting 7,310 rules into a taxonomy of 5 main and 25 sub-categories via open coding. Phase 3 details the rule evolution analysis, extracting 1,540 evolved rules, which then branch into reason analysis (yielding 504 clear rules and 6 driving reasons) and compliance assessment (evaluating 275 verifiable rules). The Survey track spans the next three phases: Phase 4 outlines the questionnaire design, including draft design and pilot surveys. Phase 5 illustrates data collection via GitHub email extraction and localized dispatch, yielding 117 initial responses. Phase 6 involves quality filtering, resulting in 99 valid responses. Finally, both the mining and survey tracks converge into a ``Synthesis \& Triangulation'' step on the far right. This step combines objective mining data with subjective developer perceptions to achieve a comprehensive understanding of rule taxonomy and evolution.}
\label{fig:study_design_overview}
\end{figure*}

Our study uses a mixed-methods approach consisting of two main phases: (1) analyzing rules collected from open-source projects developed using AI IDEs, and (2) surveying practitioners about their rule usage practices. First, we collected rule files from relevant projects by mining GitHub repositories to answer RQ1 regarding the categories of rules present in these projects. Subsequently, we conducted a practitioner survey to validate the mining results and explore how these rules evolve during development (RQ2). This methodological triangulation~\cite{Se1999} strengthens the findings by integrating objective mining data with developers' perspectives. Figure \ref{fig:study_design_overview} shows an overview of the research methodology employed in this study.

\subsection{Objective and Research Questions}\label{sec_RQ}
Our study aims to empirically investigate the practices of configuring and maintaining rules in projects developed using AI IDEs. We examine the categories of explicitly defined rules used to govern the behavior of AI IDEs and explore how these rules evolve over time. This includes identifying the reasons for rule evolution and evaluating how the compliance of software artifacts changes after these rule updates. Specifically, we address the following Research Questions (RQs):

\textbf{RQ1:} \textit{What are the categories of rules in OSS projects developed by AI IDEs?}

\textbf{Motivation.} Although rules are an emerging software artifact in AI IDEs, their application in software development is not yet well understood. RQ1 addresses this by classifying these rules through an empirical study. Constructing a taxonomy provides a structured view of how developers use rules to regulate AI IDE generated projects. Additionally, this taxonomy establishes the basis for our subsequent analysis of evolutionary trends across different rule categories.

\textbf{RQ2:} \textit{How do rules in OSS projects developed by AI IDEs evolve?}

\textbf{Motivation.} Rules change throughout a software project's lifecycle. Understanding these evolutionary characteristics is necessary for evaluating the maintenance and evolution effort required for AI-assisted development. We decompose RQ2 into three sub-RQs.

\textbf{RQ2.1:} \textit{What categories of rules undergo evolution?}

\textbf{Motivation.} This RQ maps the changed rules to the taxonomy established in RQ1. By analyzing the distribution of evolved rules across categories and their change types (i.e., additions, modifications, and deletions), we identify which categories of rules developers update most frequently. This analysis describes the current state of rule maintenance and evolution in AI IDEs.

\textbf{RQ2.2:} \textit{What are the driving reasons for rule evolution?}

\textbf{Motivation.} Identifying the underlying reasons for rule changes explains how developers adapt rules in software development with AI IDEs. By analyzing commit messages, rule content diffs, and co-changed files, we categorize the drivers of rule evolution, such as fixing AI hallucinations, improving prompt clarity, andsynchronizing with technology stack updates. The answer of this RQ can help developers write more maintainable rules and inform AI IDE designers on how to improve rule management mechanisms.

\textbf{RQ2.3:} \textit{How does the compliance rate of software artifacts change following rule evolution?}

\textbf{Motivation.} The primary purpose of maintaining and evolving rules is to align software artifacts with developers' intent. However, it is unclear whether rule evolution improves this alignment in practice. This RQ examines how the compliance rate of software artifacts changes after rule updates, providing empirical evidence on the effectiveness of rule maintenance and evolution. Understanding these changes helps evaluate whether updating rules improves the consistency of software artifacts and informs practitioners about the practical impact of rule-driven workflows.

\subsection{Mining Study Design}\label{mining_study_design}

\subsubsection{Data Collection and Filtering}\label{mining_data_collection_filtering}
We mined open-source projects containing rule configuration files from GitHub. To ensure the consistency of our longitudinal analysis, we restricted our data collection to the \texttt{main/master} branch of each repository, with a strict cutoff date for commit histories set to 24 October 2025. The data collection and filtering process was conducted in the following three steps:

\textbf{(1) Initial Repository Search via Time-Slicing.} 
We selected five AI IDEs that support rule mechanisms (Cursor, Windsurf, Trae, Qoder, and Kiro). We queried the GitHub Repository Search API\footnote{\url{https://api.github.com/search/repositories}} to identify projects whose \texttt{description} or \texttt{README} files matched specific inclusion keywords (see Table \ref{tab:keywords}). For instance, when searching for Cursor projects, we used keywords like ``\textit{cursor IDE}'' and ``\textit{by cursor}''. Because the GitHub Search API imposes a limit of 1,000 results per query, we addressed this constraint by slicing the search window chronologically. For each AI IDE, the window spanned from its release date (listed in Table \ref{tab:ai_ides}) to 15 September 2025. This time-slicing approach enabled us to collect a broader set of relevant projects within the search period.

\textbf{(2) Keyword and Rule File Filtering.} 
Because some AI IDE names are ambiguous (e.g., ``cursor'' often refers to a database cursor or a mouse pointer), we applied a set of exclusion keywords (listed in Table \ref{tab:keywords}) to remove tutorials, toy projects (e.g., \textit{test}, \textit{demo}), and repositories related to these ambiguous terms. Next, we queried the GitHub Repository Contents API using the template\footnote{\url{https://api.github.com/repos/<owner>/<repo>/contents/<rule_file_path>}} to verify the presence of the target AI IDE's rule files or directories in the remaining repositories, using the paths defined in Table \ref{tab:ai_ides}. We excluded any projects that lacked these rule files. Finally, to filter out inactive repositories, we excluded projects with fewer than 10 commits and 0 stars.

\textbf{(3) Manual Inspection.} 
The first author manually reviewed the remaining projects. This step served two purposes: first, to remove any irrelevant or toy projects that were not excluded by the keyword filtering; and second, to verify that the project's README or description explicitly stated it was developed using an AI IDE (e.g., ``\textit{This project was fully created and is maintained by AI agents}''\footnote{\url{https://github.com/petermodzelewski/yt-database}}).

Following this process, we obtained a final dataset of 83 projects (the complete project list is available in our online replication package \cite{dataset}). Table \ref{tab:filtering_results} presents a breakdown of the project counts for each AI IDE across the different filtering stages.

\begin{table*}[htbp]
\centering
\caption{Inclusion and Exclusion Keywords for Repository Search}
\scalebox{0.9}{
\label{tab:keywords}
\begin{tabular}{@{}p{0.15\linewidth} p{0.9\linewidth}@{}}
\toprule
\textbf{Category} & \textbf{Keywords} \\ \midrule
\textbf{Inclusion} \newline (e.g., Cursor) & \texttt{cursor IDE}, \texttt{cursor AI}, \texttt{using cursor}, \texttt{cursor Agent}, \texttt{use cursor}, \texttt{with cursor}, \texttt{by cursor}, \texttt{in cursor}, \texttt{through cursor}, \texttt{via cursor} \\ \midrule
\textbf{Exclusion} & \texttt{test}, \texttt{demo}, \texttt{learn}, \texttt{practice}, \texttt{practical}, \texttt{rule}, \texttt{rules}, \texttt{mouse}, \texttt{pagination}, \texttt{a cursor}, \texttt{curse}, \texttt{3D cursor}, \texttt{your cursor}, \texttt{custom cursor}, \texttt{cursor navigation}, \texttt{navigation}, \texttt{cursor move}, \texttt{example}, \texttt{awesome}, \texttt{movement}, \texttt{keyboard}, \texttt{custom}, \texttt{position}, \texttt{animat}, \texttt{pointer}, \texttt{theme}, \texttt{attempt}, \texttt{moving}, \texttt{move}, \texttt{Paginate}, \texttt{Trying}, \texttt{try}, \texttt{simple}, \texttt{quick}, \texttt{oracle}, \texttt{store procedure}, \texttt{mysql}, \texttt{sql server}, \texttt{cursor library}, \texttt{Drawing}, \texttt{draw}, \texttt{prompt}, \texttt{collection}, \texttt{hand gesture}, \texttt{canvas}, \texttt{click}, \texttt{cursor array}, \texttt{SQL}, \texttt{experiment}, \texttt{scrollbar}, \texttt{Portfolio}, \texttt{template}, \texttt{course}, \texttt{tutorial}, \texttt{toy}, \texttt{live cursor}, \texttt{guideline}, \texttt{quiz}, \texttt{small}, \texttt{exploring}, \texttt{realtime}, \texttt{real-time}, \texttt{window}, \texttt{first}, \texttt{homework} \\ \bottomrule
\end{tabular}
}
\end{table*}

\begin{table}[htbp]
\centering
\caption{Project Counts Across Data Filtering Stages}
\scalebox{0.9}{
\label{tab:filtering_results}
\begin{tabular}{@{}lrrr@{}}
\toprule
\textbf{AI IDE} & \textbf{Initial Search} & \textbf{Keyword and Rule File Filtering} & \textbf{Manual Inspection} \\ \midrule
Cursor   & 18,790 & 132 & 46 \\
Windsurf & 3,684  & 9   & 5  \\
Qoder    & 125    & 12  & 1  \\
Trae     & 1,445  & 152 & 6  \\
Kiro     & 1,543  & 189 & 25 \\ 
\textbf{Total}    & \textbf{25,587} & \textbf{494} & \textbf{83} \\ \bottomrule
\end{tabular}
}
\end{table}

\subsubsection{Rule Extraction and Taxonomy Construction}\label{mining_rule_extraction_taxonomy}
To answer RQ1, we first extracted and segmented the rules from the 83 collected projects. By querying the GitHub Repository Contents API, we extracted standalone rule files (e.g., \texttt{.cursorrules} and \texttt{.windsurfrules}) as well as all files residing within specific rule directories (e.g., \texttt{.cursor/rules/}). Most of these artifacts are formatted as Markdown files. The AI IDE Kiro automatically generates three default files under its \texttt{.kiro/steering/} directory: \texttt{product.md}, \texttt{tech.md}, and \texttt{structure.md}. Because this study focuses on rules that guide and constrain AI IDE coding behaviors, technical specifications, and collaboration patterns, we excluded \texttt{product.md}. This file primarily documents business requirements and product features, which deviate from our focus on development constraints. Following this extraction phase, we obtained 325 valid rule files from the 83 projects.

\begin{figure}[htbp]
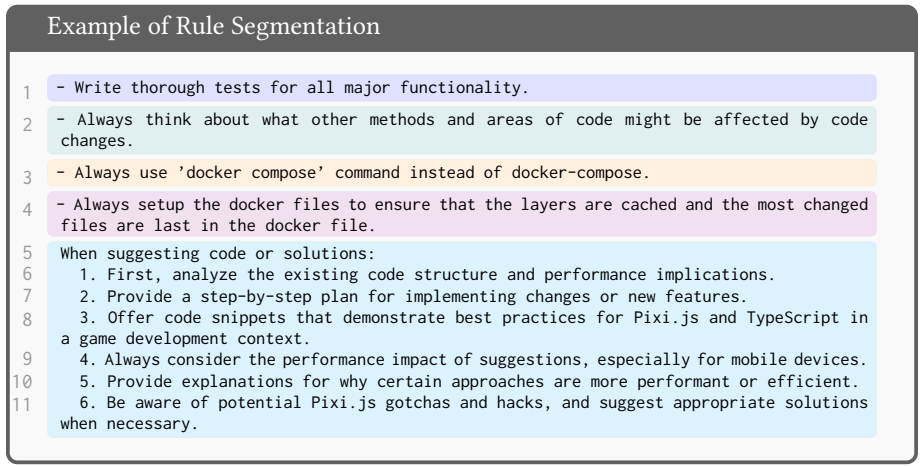

    \centering

    \begin{tcolorbox}[colback=gray!5!white,colframe=gray!75!black,title=Example of Rule Segmentation]
    
    \begin{tcolorbox}[
            enhanced,
            colback=blue!12,
            colframe=blue!20,
            boxrule=0pt,
            arc=2pt,
            boxsep=0pt,
            left=5pt, right=3pt, top=2pt, bottom=2pt,
            fontupper=\scriptsize\ttfamily,
            before skip=2pt,
            after skip=2pt,
            before upper=\parindent0pt,
            overlay={
                \node[anchor=north east, text=gray!80, font=\footnotesize\ttfamily]
                at ([xshift=-2pt, yshift=-0.5pt]frame.north west) {1};
            }
        ]
- Write thorough tests for all major functionality.
    \end{tcolorbox}
    
    \begin{tcolorbox}[
            enhanced,
            colback=teal!12,
            colframe=gray!20,
            boxrule=0pt,
            arc=2pt,
            boxsep=0pt,
            left=5pt, right=3pt, top=2pt, bottom=2pt,
            fontupper=\scriptsize\ttfamily,
            before skip=2pt,
            after skip=2pt,
            before upper=\parindent0pt,
            overlay={
                \node[anchor=north east, text=gray!80, font=\footnotesize\ttfamily]
                at ([xshift=-2pt, yshift=-0.5pt]frame.north west) {2};
            }
        ]
- Always think about what other methods and areas of code might be affected by code changes.
    \end{tcolorbox}
    
    \begin{tcolorbox}[
            enhanced,
            colback=orange!12,
            colframe=gray!20,
            boxrule=0pt,
            arc=2pt,
            boxsep=0pt,
            left=5pt, right=3pt, top=2pt, bottom=2pt,
            fontupper=\scriptsize\ttfamily,
            before skip=2pt,
            after skip=2pt,
            before upper=\parindent0pt,
            overlay={
                \node[anchor=north east, text=gray!80, font=\footnotesize\ttfamily]
                at ([xshift=-2pt, yshift=-0.5pt]frame.north west) {3};
            }
        ]
- Always use 'docker compose' command instead of docker-compose.
    \end{tcolorbox}
    
    \begin{tcolorbox}[
            enhanced,
            colback=violet!12,
            colframe=gray!20,
            boxrule=0pt,
            arc=2pt,
            boxsep=0pt,
            left=5pt, right=3pt, top=2pt, bottom=2pt,
            fontupper=\scriptsize\ttfamily,
            before skip=2pt,
            after skip=2pt,
            before upper=\parindent0pt,
            overlay={
                \node[anchor=north east, text=gray!80, font=\footnotesize\ttfamily]
                at ([xshift=-2pt, yshift=-0.5pt]frame.north west) {4};
            }
        ]
- Always setup the docker files to ensure that the layers are cached and the most changed files are last in the docker file.
    \end{tcolorbox}
    
    \begin{tcolorbox}[
            enhanced,
            colback=cyan!12,
            colframe=gray!20,
            boxrule=0pt,
            arc=2pt,
            boxsep=0pt,
            left=5pt, right=3pt, top=2pt, bottom=2pt,
            fontupper=\scriptsize\ttfamily,
            before skip=2pt,
            after skip=2pt,
            before upper=\parindent0pt,
            overlay={
                \node[anchor=north east, text=gray!80, font=\footnotesize\ttfamily] at ([xshift=-2pt, yshift=2pt]frame.north west) {5};
                \node[anchor=north east, text=gray!80, font=\footnotesize\ttfamily] at ([xshift=-2pt, yshift=-6pt]frame.north west) {6};
                \node[anchor=north east, text=gray!80, font=\footnotesize\ttfamily] at ([xshift=-2pt, yshift=-14pt]frame.north west) {7};
                \node[anchor=north east, text=gray!80, font=\footnotesize\ttfamily] at ([xshift=-2pt, yshift=-23pt]frame.north west) {8};
                \node[anchor=north east, text=gray!80, font=\footnotesize\ttfamily] at ([xshift=-2pt, yshift=-38pt]frame.north west) {9};
                \node[anchor=north east, text=gray!80, font=\footnotesize\ttfamily] at ([xshift=-2pt, yshift=-46.5pt]frame.north west) {10};
                \node[anchor=north east, text=gray!80, font=\footnotesize\ttfamily] at ([xshift=-2pt, yshift=-55pt]frame.north west) {11};
            }
        ]
When suggesting code or solutions:\\ 
\hspace*{1em}1. First, analyze the existing code structure and performance implications.\\
\hspace*{1em}2. Provide a step-by-step plan for implementing changes or new features.\\
\hspace*{1em}3. Offer code snippets that demonstrate best practices for Pixi.js and TypeScript in a game development context.\\
\hspace*{1em}4. Always consider the performance impact of suggestions, especially for mobile devices.\\
\hspace*{1em}5. Provide explanations for why certain approaches are more performant or efficient.\\
\hspace*{1em}6. Be aware of potential Pixi.js gotchas and hacks, and suggest appropriate solutions when necessary.
    \end{tcolorbox}
    
    \end{tcolorbox} 
    
    \caption{An example of segmenting rule text into distinct semantic units (rules).}
    \Description{A formatted example illustrating how a continuous block of text in a rule file is segmented into individual semantic units. Lines 1 through 4 are highlighted as separate, independent rules with gaps between them, whereas lines 5 through 11 are grouped together in a single contiguous block to represent one complex, multi-line rule.}
    \label{fig:rule_segmentation_example}
\end{figure}

Because a single rule file typically contains multiple independent rules, we established ``a single rule'' as our fundamental unit of analysis. The first author manually reviewed and segmented all files. To ensure the reliability of this segmentation process, the first and third authors resolved ambiguous boundaries through discussion, while the other co-authors audited the results and provided feedback. We defined a complete ``semantic unit'' as the smallest indivisible text block that conveys a standalone, actionable constraint for the AI IDE. The boundary of a rule is determined by its contextual independence. In most cases, a single line of text (e.g., an independent bullet point) encapsulates a complete constraint and is therefore treated as a single rule. However, multiple lines are merged into a single cohesive rule if they exhibit structural or logical dependencies that render them incomplete or meaningless if separated. For instance, we preserved multi-line blocks as single units when they comprised a conditional trigger followed by a nested list of steps, or a high-level directive accompanied by explanatory code snippets. As illustrated in Figure \ref{fig:rule_segmentation_example}, Lines 1, 2, 3, and 4 each represent an independent rule because they are contextually self-contained. Each color corresponds to one complete rule. In contrast, Lines 5 through 11 collectively form a single rule because the numbered steps (Lines 6--11) are logically bound to the conditional prerequisite stated in Line 5 (``When suggesting code or solutions:''). Separating these blocks would cause the individual steps to lose their triggering condition.

After completing the segmentation, we extracted 7,310 individual rules from the 325 files. We then employed an open coding approach \cite{Se1999} to categorize these rules (the raw files and coding mapping datasets are publicly available in our replication package \cite{dataset}). The first author reviewed each rule and assigned an initial code. When disagreements or uncertainties arose, the first and third authors held discussions and applied the negotiated agreement approach \cite{CaQuOsPe2013} to reach a consensus. For example, referring back to Figure \ref{fig:rule_segmentation_example}, we coded Lines 1 through 4 as \textit{Testing Strategy}, \textit{Design Principles}, \textit{Environment Configuration}, and \textit{Environment Configuration}, respectively. Conversely, we coded the entire block from Lines 5 to 11 as \textit{AI Behavior \& Decision Strategies}. Through this process of induction and abstraction, we constructed a hierarchical taxonomy comprising 5 primary categories and 25 secondary categories.

\subsubsection{Evolution Rule Extraction and Categorization}\label{mining_evolution_rule_taxonomy}
We focused on the \texttt{main/master} branch of each repository and extracted the evolutionary history of the 325 rule files collected in Section \ref{mining_rule_extraction_taxonomy}. To account for file renames or relocations during project iterations, we used Git commands to track the complete lifecycle path of each rule file. We extracted historical commit hashes, timestamps, and corresponding code differences (Diffs) to create chronologically ordered evolution sequences. Statistics show that 55 of the 83 projects contain rule files that underwent evolution. At the file level, 117 of the 325 rule files contain at least one modification (i.e., commit count > 1).

\begin{figure}[htbp]
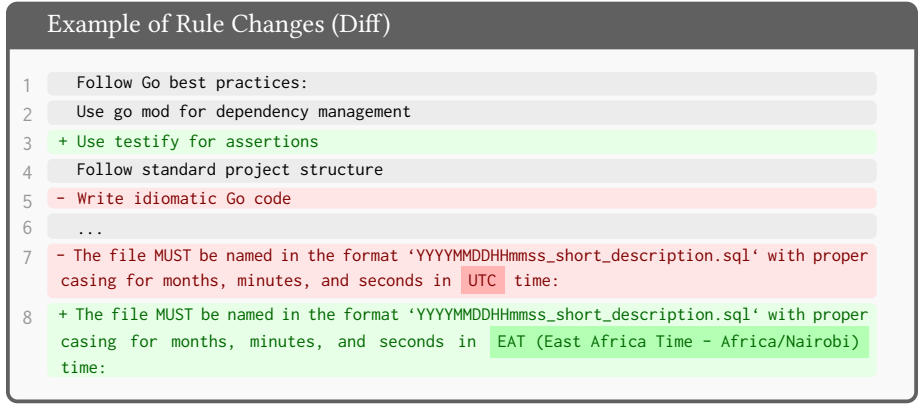

    \centering
    \begin{tcolorbox}[colback=gray!5!white,colframe=gray!75!black,title=Example of Rule Changes (Diff)]
    
    \tcbset{
        diffbox/.style={
            enhanced,
            boxrule=0pt,
            arc=2pt,
            boxsep=0pt,
            left=5pt, right=3pt, top=1.5pt, bottom=1.5pt,
            fontupper=\scriptsize\ttfamily\linespread{0.92}\selectfont,
            before skip=2pt, after skip=2pt,
            before upper=\parindent0pt,
            overlay={
                \node[anchor=north east, text=gray!80, font=\footnotesize\ttfamily] 
                at ([xshift=-2pt, yshift=1pt]frame.north west) {#1};
            }
        },
        diffgray/.style={diffbox=#1, colback=gray!15, colframe=gray!15},
        diffadd/.style={diffbox=#1, colback=green!10, colframe=green!10, coltext=green!40!black},
        diffdel/.style={diffbox=#1, colback=red!10, colframe=red!10, coltext=red!50!black}
    }
    
    \begin{tcolorbox}[diffgray=1]
  \hspace{0.85em}Follow Go best practices:
    \end{tcolorbox}
    
    \begin{tcolorbox}[diffgray=2]
    \hspace{0.85em}Use go mod for dependency management
    \end{tcolorbox}
    
    \begin{tcolorbox}[diffadd=3]
+   Use testify for assertions
    \end{tcolorbox}
    
    \begin{tcolorbox}[diffgray=4]
    \hspace{0.85em}Follow standard project structure
    \end{tcolorbox}
    
    \begin{tcolorbox}[diffdel=5]
-   \hspace{0.15em}Write idiomatic Go code
    \end{tcolorbox}
    
    \begin{tcolorbox}[diffgray=6]
    \vspace{0.8em} \hspace{0.85em}...
    \end{tcolorbox}
    
    \begin{tcolorbox}[diffdel=7]
- The file MUST be named in the format `YYYYMMDDHHmmss\_short\_description.sql` with proper casing for months, minutes, and seconds in \colorbox{red!30}{UTC} time:
    \end{tcolorbox}
    
    \begin{tcolorbox}[diffadd=8]
+ The file MUST be named in the format `YYYYMMDDHHmmss\_short\_description.sql` with proper casing for months, minutes, and seconds in \colorbox{green!30}{EAT (East Africa Time - Africa/Nairobi)} time:
    \end{tcolorbox}
    
    \end{tcolorbox} 
    
    \caption{An example of extracting changes from rule Diffs.}
    \Description{A formatted example illustrating Git Diff representation of rule evolution. It shows context lines in gray, an added rule in light green with a '+' sign, a deleted rule in light red with a '-' sign, and a modified rule represented by a deleted line followed by an added line. Specific changed words in the modified rule (``UTC'' changed to ``EAT'') are highlighted with darker background colors to pinpoint the exact modification.}
    \label{fig:rule_diff_example}
\end{figure}

We performed fine-grained rule segmentation on the extracted Diffs using the same criteria described in Section \ref{mining_rule_extraction_taxonomy}. Concurrently, we labeled the change type of each rule as \textit{Added}, \textit{Modified}, or \textit{Deleted}. Figure \ref{fig:rule_diff_example} illustrates these types: Line 3 represents an added rule, Line 5 represents a deleted rule, and Lines 7 and 8 demonstrate a modified rule. We identified and excluded trivial changes limited to punctuation, whitespace, or line breaks (i.e., \textit{Chore} modifications).

Finally, we employed a deductive coding approach \cite{FeMu2006} to categorize the evolved rules. The first author used the two-level taxonomy from Section \ref{mining_rule_extraction_taxonomy} as a codebook to map each evolved rule to a category. We resolved ambiguous mappings through offline discussions between the first and third authors. We extracted and segmented 1,894 evolved rules from the Diffs of the 117 files. After filtering out 354 \textit{Chore} modifications, we retained 1,540 valid evolved rules for subsequent analysis (the raw Diff blocks and categorized mapping dataset are available in our replication package \cite{dataset}).

\subsubsection{Analysis of Reasons for Rule Evolution}\label{mining_reason_for_rules}
\begin{figure*}[t]
\centering
\includegraphics[
    width=\textwidth,
]{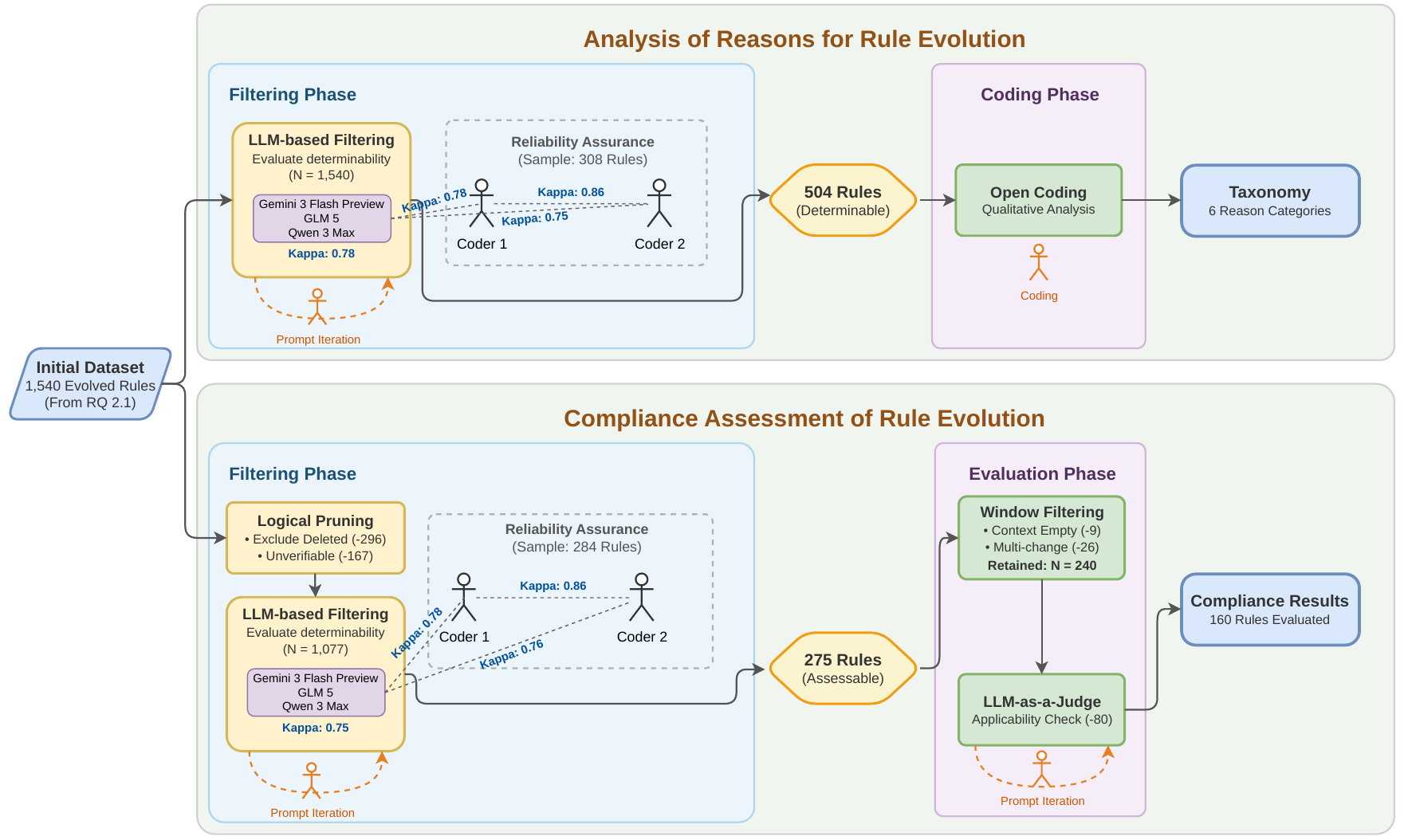}
\caption{Overview of the rule evolution analysis and compliance assessment process.}
\Description{This flowchart illustrates the methodological workflow for evaluating rule evolution, divided into two parallel horizontal branches: ``Analysis of Reasons'' (Section \ref{mining_reason_for_rules}) at the top, and ``Compliance Assessment'' (Section \ref{mining_compliance_of_rule}) at the bottom. Both branches begin with a shared initial dataset of 1,540 evolved rules.  The top branch features a Filtering Phase, where LLMs filter the 1,540 rules down to 504 determinable rules. This is supported by a Reliability Assurance check that calculates Kappa scores among human coders and three different LLMs. The rules then pass to a Coding Phase, which uses open coding to output a final taxonomy of 6 reason categories. The bottom branch also starts with a Filtering Phase, which prunes and LLM-filters the rules down to 275 assessable rules, verified by a similar Reliability Assurance check. The rules then enter the Evaluation Phase, undergoing window filtering and an LLM-as-a-judge applicability check to output final compliance results for 160 evaluated rules. In both branches, the LLM evaluation steps explicitly depict a prompt iteration loop conducted by a researcher.}
\label{fig:reason_and_compliance_procedure}
\end{figure*}

After extracting the 1,540 valid evolved rules in Section \ref{mining_evolution_rule_taxonomy}, we investigated the reasons behind these changes by analyzing their contextual artifacts, including rule content, change types, commit messages, and co-changed files. Figure \ref{fig:reason_and_compliance_procedure} outlines the overall procedure for this analysis and the subsequent compliance assessment. Because rules and changes vary in granularity, not every rule evolution provides sufficient context to determine its cause (e.g., changes with uninformative commit messages). Therefore, we first filtered the 1,540 rules based on determinability, assessing whether the rationale for each change could be objectively identified.

Because manually reviewing 1,540 rules is labor-intensive, and recent studies have demonstrated the feasibility of applying LLMs to mining software repositories \cite{DeCaPaFrMa2025, CoLaNo2025, CoLaNoQu2024}, we used LLMs for automated data labeling, complemented by human sampling to verify reliability. Based on Cochran's sample size formula \cite{Ah2024}, a population of 1,540 requires a minimum sample of 308 to achieve a 95\% confidence level with a 5\% margin of error. We randomly sampled 308 rules to establish a human-labeled ground truth. The first author and a software engineering practitioner independently conducted a pilot labeling on 30 initial samples, resulting in a Cohen's Kappa of 0.76. After resolving disagreements and aligning their criteria, the two human coders independently labeled the full 308-rule subset. The final Cohen's kappa was 0.86, indicating strong inter-rater agreement.

Using the insights from the human labeling process, we iteratively refined prompts to guide the LLMs in assessing the determinability of 1,540 rules. To reduce potential model bias, we independently applied three LLMs (Gemini 3 Flash Preview, GLM 5, and Qwen3 Max). During the prompt refinement, we analyzed cases where the LLMs disagreed with each other or deviated from the human ground truth, and we updated the instructions to handle these edge cases. Figure \ref{fig:prompt_determinability} shows the final prompt used for the LLMs (with minor omissions for brevity; the full version is available in our replication package \cite{dataset}). This prompt contains an Ultra-Conservative inference protocol, which mandates that the models adhere to the principle of Evidence over Inference. Specifically, a rule change is classified as \textit{Clear} only if there is a structural match between the rule difference and the co-changed file paths, or a semantic alignment with the commit message (e.g., a commit message stating ``\textit{refactor(rules): update instruction compliance and conventional commit rules for better enforcement}''). Otherwise, vague or generic changes are classified as \textit{Unclear}. For the full dataset, the Fleiss' Kappa~\cite{FaQu2015} among the three LLMs was 0.78. For the 308-rule sample, the Cohen's Kappa~\cite{cohen1960coefficient} between the LLMs' majority vote and the two human coders was 0.78 and 0.75, respectively. These metrics demonstrate consistency both among the LLMs and between the LLMs and human evaluators.

\begin{figure}[htbp]
    \centering
    \begin{tcolorbox}[
        colback=orange!3!white,      
        colframe=orange!80!black,    
        title=Abridged Prompt for Reason Determinability Filtering,
        fonttitle=\bfseries\small,
        boxrule=0.8pt,
        arc=3pt,
        top=2pt, bottom=2pt, left=2pt, right=2pt
    ]
    
    \begin{tcolorbox}[
        enhanced,
        colback=orange!8!white,      
        colframe=orange!50!black,
        boxrule=0.5pt,
        arc=2pt,
        top=10pt, bottom=4pt, left=6pt, right=6pt, 
        fontupper=\scriptsize\ttfamily\linespread{1.1}\selectfont\raggedright, 
        title={\sffamily\bfseries System Prompt},
        attach boxed title to top left={xshift=10pt, yshift=-8pt},
        boxed title style={colback=orange!85!black, colframe=orange!85!black, size=small}
    ]
\# Role: You are an ultra-conservative Researcher in Software Evolution. Your task is to analyze a specific **change event** of an AI IDE Rule and determine the **Primary Driver (Reason)**.

\par\vspace{3pt}

\# Core Philosophy: "Evidence over Inference."
You are strictly biased towards "Unclear". You only assign a specific category if there is **DIRECT, UNDENIABLE evidence** linking the Rule Diff to the Context. You must explicitly punish generic advice, unannotated deletions, and speculative causality.

\par\vspace{3pt}
\textcolor{gray}{...(Omitted content)}
\par\vspace{3pt}

\# Inference Logic (The Ultra-Conservative Protocol)

\par\vspace{3pt}
\textcolor{gray}{...(Omitted content)}
\par\vspace{3pt}

**Step 1: The "Direct Link" Check (Structural or Intentional)**
\par
*   **Structural Match (Absolute Proof):** Is there an **exact keyword, precise filename, or unique path match** between the Rule text and the Co-changed Files list (e.g., Rule mentions `extractors/` and Co-changed files include `src/extractors/`)?
\par
    *   *YES:* This is definitive evidence of Synchronization or Context Enrichment. **This structural evidence overrides a seemingly unrelated commit message.**
\par
*   **Intentional Match:** Does the Commit Message perfectly and explicitly describe the exact change made in the Rule Diff?
\par
    *   *YES:* Strong evidence for a specific category.
\par
*   *NO to both:* Do NOT assume a link just because they use the same framework/language. Proceed to Step 2.

**Step 2
\par\vspace{3pt}
\textcolor{gray}{...(Omitted content)}
\par\vspace{3pt}

\# Output Format
Return a SINGLE JSON object: 
\par
\{
\par
"tier\_1": "'Clear' or 'Unclear'", 
\par
"confidence": "...", 
\par
"reasoning": "A single sentence. If Unclear, state why the evidence is insufficient."
\par
\}
\par\vspace{3pt}
\# Reference Examples (Learn the Decision Boundary)
\par\vspace{3pt}
\textcolor{gray}{...(Omitted content)}
\par\vspace{3pt}
    \end{tcolorbox}
    
    \vspace{2pt}

    \begin{tcolorbox}[
        enhanced,
        colback=orange!1!white,             
        colframe=orange!50!black,
        boxrule=0.5pt,
        arc=2pt,
        top=10pt, bottom=4pt, left=6pt, right=6pt, 
        fontupper=\scriptsize\ttfamily\linespread{1.1}\selectfont\raggedright, 
        title={\sffamily\bfseries User Prompt},
        attach boxed title to top left={xshift=10pt, yshift=-8pt},
        boxed title style={colback=orange!85!black, colframe=orange!85!black, size=small}
    ]
\# Input Data for Analysis
\par
\#\# 1. Rule Info
\par
* **Project Name:** \{project\_name\}
\par
* **Target Rule File:** \{rule\_file\_path\}
\par
* **Change Type:** \{change\_type\}
\par
* **Rule Content (Diff):** \{rule\_content\}
\par\vspace{3pt}
\#\# 2. Commit Info
\par
* **Commit Message:** \{commit\_message\}
\par\vspace{3pt}
\#\# 3. Co-changed Files (Format: change\_type file\_path added\_lines deleted\_lines)
\par
\{co\_changed\_files\_raw\_list\}
\par\vspace{3pt}
\# Task
Based on the inputs above, categorize the driver of this rule change.
    \end{tcolorbox}

    \end{tcolorbox}
    \caption{The prompt structure used for LLM-based reason determinability filtering.}
    \Description{A block diagram showing the structure of the LLM prompt in a professional blue theme. It is divided into a 'System Prompt' box detailing the conservative inference logic, and a white 'User Prompt' box defining the data structure for project names, rule differences, commit messages, and co-changed files.}
    \label{fig:prompt_determinability}
\end{figure}

Following this LLM cross-filtering, we extracted a subset of 504 rules that all three models unanimously determined as \textit{Clear}. The first author then applied open coding to these 504 rules to identify the reasons for their evolution, resolving boundary ambiguities through discussions with the third author. During this process, we developed guiding criteria to support the classification of the reasons for rule evolution. For example, if a rule addition accompanied the creation of new files, we distinguished between \textit{Context Enrichment} (providing static documentation or references) and \textit{Expansion} (imposing new logical constraints). We classified a rule modification synchronized with a file rename or move as \textit{Synchronization}. We categorized changes that introduced negative constraints (e.g., ``Do NoT'') or were associated with fix-related commit messages as \textit{Correction}, and we classified non-logical adjustments that improved tone or formatting as \textit{Refinement}. Through this open coding, we inductively derived 6 primary categories of reasons for rule evolution.

\subsubsection{Compliance Assessment of Rule Evolution}\label{mining_compliance_of_rule}
To investigate the practical impact of rule evolution, we conducted a longitudinal assessment of the evolved rules to measure how the compliance of software artifacts (e.g., code, configuration files, documentation, directory structures, and commit messages) changed before and after rule modifications. As illustrated in Figure \ref{fig:reason_and_compliance_procedure}, this pipeline consists of three phases: preliminary exclusion, verifiability filtering, and compliance evaluation via LLM-as-a-judge \cite{HeShZhTrSuXiDuLo2025}.

First, during the preprocessing phase, we pruned the initial 1,540 evolved rules. We excluded 296 rules with a \textit{deleted} change type because tracking compliance for a removed constraint is obsolete. We also filtered out 167 rules from categories that lack statically verifiable artifacts (e.g., soft guidelines for AI interaction behavior or subjective code review instructions without objective metrics). Following these exclusions, 1,077 rules remained for analysis.

Second, similar to the data filtering process for ambiguous rules (Section \ref{mining_reason_for_rules}), not all remaining rules could be objectively assessed for compliance via static artifacts. Because manual verification is labor-intensive, we again used LLMs for data labeling and validated the results through manual sampling. Based on Cochran's formula \cite{Ah2024}, a population of 1,077 requires a minimum sample of 284 to achieve a 95\% confidence level with a 5\% margin of error. Therefore, we randomly selected 284 rules for manual labeling. After resolving discrepancies, the Cohen's Kappa~\cite{cohen1960coefficient} between the first author and another software engineering practitioner reached 0.86. We then independently applied three LLMs (Gemini 3 Flash Preview, GLM 5, and Qwen3 Max) to evaluate the entire dataset. Figure \ref{fig:prompt_verifiability} shows the abridged prompt for this task (the full version is available in our replication package \cite{dataset}). In this prompt, we applied a static verifiability criterion: a rule is verifiable only if its compliance can theoretically be checked by a simple script (e.g., regular expressions, AST parsing) without false positives. We excluded rules requiring runtime execution, complex control flow tracing, or subjective intent interpretation, as well as rules serving merely as references. Conversely, rules specifying explicit file paths or naming conventions were considered verifiable. The Fleiss' Kappa \cite{FaQu2015} among the three LLMs was 0.75, and the Cohen's Kappa between the LLMs' majority vote and the two human annotators was 0.78 and 0.76, respectively. Following this step, we extracted 275 rules that all three models unanimously identified as statically verifiable.

\begin{figure}[htbp]
    \centering
    \begin{tcolorbox}[
        colback=orange!3!white,      
        colframe=orange!80!black,    
        title=Abridged Prompt for Rule Verifiability Filtering,
        fonttitle=\bfseries\small,
        boxrule=0.8pt,
        arc=3pt,
        top=2pt, bottom=2pt, left=2pt, right=2pt
    ]
    
    \begin{tcolorbox}[
        enhanced,
        colback=orange!8!white,      
        colframe=orange!50!black,
        boxrule=0.5pt,
        arc=2pt,
        top=10pt, bottom=4pt, left=6pt, right=6pt, 
        fontupper=\scriptsize\ttfamily\linespread{1.1}\selectfont\raggedright, 
        title={\sffamily\bfseries System Prompt},
        attach boxed title to top left={xshift=10pt, yshift=-8pt},
        boxed title style={colback=orange!85!black, colframe=orange!85!black, size=small}
    ]
\# Role: You are an expert MSR (Mining Software Repositories) Researcher and Automated Testing Architect. Your task is to analyze "AI IDE Rules" and determine if they are **Technically Verifiable** based on concrete software artifacts.

\par\vspace{3pt}
\textcolor{gray}{...(Omitted content)}
\par\vspace{3pt}

\# The Core Question (The "Simple Script" Golden Rule)
\par
**"Can I write a simple, deterministic script (using ONLY simple Regex, basic AST parsing, or file path checks) to verify this rule across thousands of commits WITHOUT ANY false positives/negatives?"
\par
**If the answer requires "understanding developer intent", "tracing complex control flows", or "verifying algorithmic correctness", the answer is unequivocally FALSE.

\par\vspace{3pt}

\# 1. Strictly Permitted Static Targets (Must be tracked in Git)
\par
To be TRUE, the rule MUST uniquely target one of the following:
\par
- Raw Source Code syntax: Exact keywords, specific annotations/decorators (e.g., `@Component`), explicit type declarations (e.g., `any`).
\par
- File System Structure: Existence of specifically named source files (e.g., `tsconfig.json`, `*route.dart`).

\par\vspace{3pt}
\textcolor{gray}{...(Omitted content)}
\par\vspace{3pt}

\par\vspace{3pt}

\# 2. STRICTLY BANNED (Will immediately result in FALSE)
\par
If the rule involves ANY of the following, you MUST mark can\_detect: false:
\par
- [BANNED] Runtime/Execution: "Tests pass", "App loads fast", "Memory usage".

\par\vspace{3pt}
\textcolor{gray}{...(Omitted content)}
\par\vspace{3pt}

\# Output Format
\par
Return a SINGLE JSON object:
\par
\{ 
\par
"can\_detect": true/false, 
\par
"detection\_targets": "...", 
\par
"detection\_logic": "..." 
\par
\}

\# Few-Shot Examples (Reference Cases)
\par\vspace{3pt}
\textcolor{gray}{...(Omitted content)}
\par\vspace{3pt}
    \end{tcolorbox}
    
    \vspace{2pt}

    \begin{tcolorbox}[
        enhanced,
        colback=orange!1!white,             
        colframe=orange!50!black,
        boxrule=0.5pt,
        arc=2pt,
        top=10pt, bottom=4pt, left=6pt, right=6pt, 
        fontupper=\scriptsize\ttfamily\linespread{1.1}\selectfont\raggedright, 
        title={\sffamily\bfseries User Prompt},
        attach boxed title to top left={xshift=10pt, yshift=-8pt},
        boxed title style={colback=orange!85!black, colframe=orange!85!black, size=small}
    ]
\# Input Rule
\par
\{
\par
  "change\_type": "\{change\_type\}", 
  \par
  "first\_level": "\{first\_level\}",
  \par
  "second\_level": "\{second\_level\}",
  \par
  "rule\_content": "\{rule\_content\}"
  \par
\}

    \end{tcolorbox}

    \end{tcolorbox}
    \caption{The prompt structure used for LLM-based verifiability filtering.}
    \Description{The figure illustrates the prompt structure used for rule verifiability filtering. The prompt consists of two parts: a system prompt and a user prompt. The system prompt defines the LLM's role as an expert in Mining Software Repositories and automated testing, and introduces a core criterion—whether a rule can be verified using a simple and deterministic script (e.g., regex matching, basic AST parsing, or file path checks) across large-scale commits without ambiguity. It further specifies strictly permitted static targets (such as source code syntax and file structures) while explicitly excluding rules that depend on runtime behavior or semantic understanding. The user prompt provides the input rule in a structured JSON format. The model is required to output a standardized JSON object containing the verifiability decision, detection targets, and detection logic, enabling automated assessment of rule verifiability.}
    \label{fig:prompt_verifiability}
\end{figure}

For these 275 rules, we defined an observation window consisting of up to 5 commits before and 5 commits after the rule change event (or the available commits if fewer than 5 existed). To reduce evaluation noise, we filtered out tangled commits and large files. The filtering criteria excluded: (1) commits modifying more than 50 files; (2) commits modifying more than 2,000 lines; (3) individual files with more than 1,000 modified lines; and (4) individual files whose total length exceeds 1,000 lines. Triggering these criteria only excluded the specific commit or file, rather than immediately discarding the rule. However, if applying these filters eliminated all preceding or all succeeding commits for a specific rule, we discarded that rule (removing 9 rules). Furthermore, to prevent analysis conflicts, we discarded rules that were subsequently modified or deleted in a later commit within the window, or had already been created or modified in a prior commit (removing 26 rules). Consequently, 240 rules remained for the LLM-as-a-judge evaluation.

\begin{figure}[htbp]
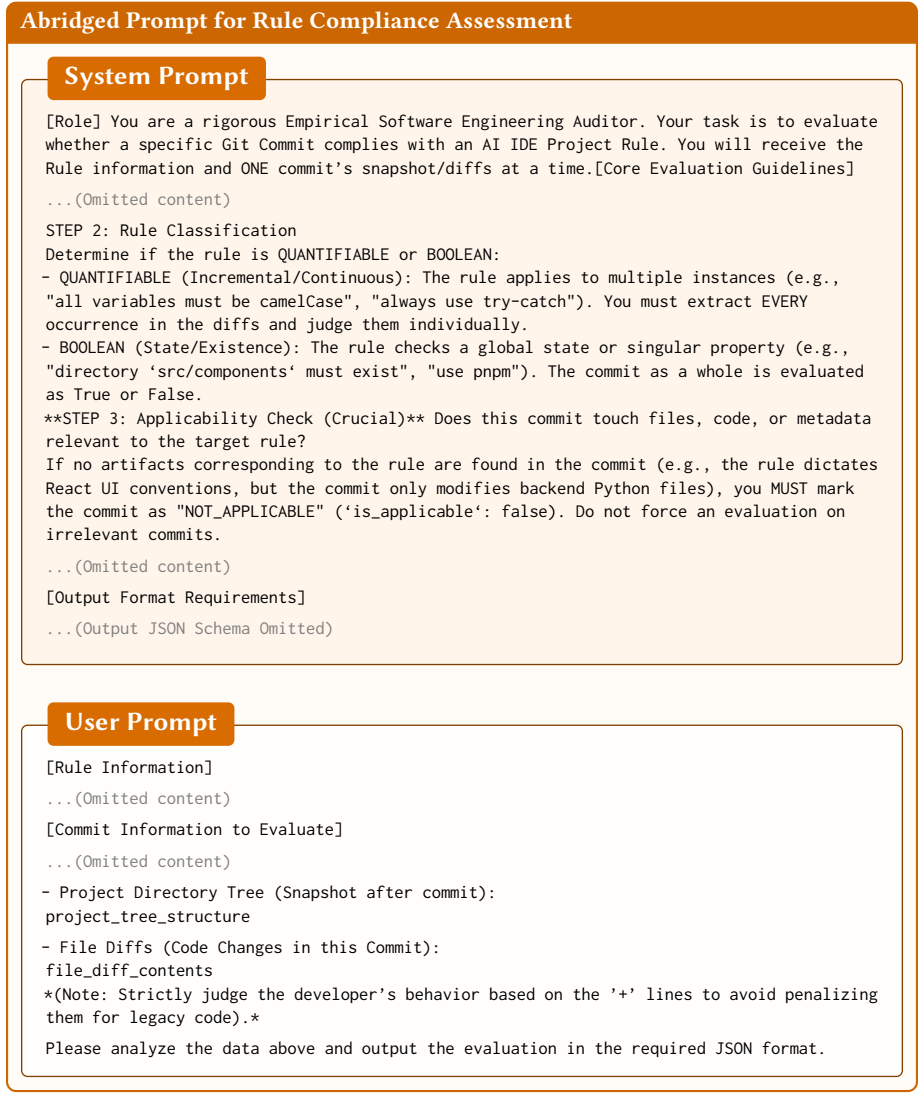

    \centering
    \begin{tcolorbox}[
        colback=orange!3!white,      
        colframe=orange!80!black,    
        title=Abridged Prompt for Rule Compliance Assessment,
        fonttitle=\bfseries\small,
        boxrule=0.8pt,
        arc=3pt,
        top=2pt, bottom=2pt, left=2pt, right=2pt
    ]
    
    \begin{tcolorbox}[
        enhanced,
        colback=orange!8!white,      
        colframe=orange!50!black,
        boxrule=0.5pt,
        arc=2pt,
        top=10pt, bottom=4pt, left=6pt, right=6pt, 
        fontupper=\scriptsize\ttfamily\linespread{1.1}\selectfont\raggedright, 
        title={\sffamily\bfseries System Prompt},
        attach boxed title to top left={xshift=10pt, yshift=-8pt},
        boxed title style={colback=orange!85!black, colframe=orange!85!black, size=small}
    ]
[Role] You are a rigorous Empirical Software Engineering Auditor. Your task is to evaluate whether a specific Git Commit complies with an AI IDE Project Rule. You will receive the Rule information and ONE commit's snapshot/diffs at a time.[Core Evaluation Guidelines]

\par\vspace{3pt}
\textcolor{gray}{...(Omitted content)}
\par\vspace{3pt}

STEP 2: Rule Classification
\par
Determine if the rule is QUANTIFIABLE or BOOLEAN:
\par
- QUANTIFIABLE (Incremental/Continuous): The rule applies to multiple instances (e.g., "all variables must be camelCase", "always use try-catch"). You must extract EVERY occurrence in the diffs and judge them individually.
\par
- BOOLEAN (State/Existence): The rule checks a global state or singular property (e.g., "directory `src/components` must exist", "use pnpm"). The commit as a whole is evaluated as True or False.

\par

**STEP 3: Applicability Check (Crucial)**
Does this commit touch files, code, or metadata relevant to the target rule?
\par
If no artifacts corresponding to the rule are found in the commit (e.g., the rule dictates React UI conventions, but the commit only modifies backend Python files), you MUST mark the commit as "NOT\_APPLICABLE" (`is\_applicable`: false). Do not force an evaluation on irrelevant commits.

\par\vspace{3pt}
\textcolor{gray}{...(Omitted content)}
\par\vspace{3pt}

[Output Format Requirements]

\par\vspace{3pt}
\textcolor{gray}{...(Output JSON Schema Omitted)}
\par\vspace{3pt}
    \end{tcolorbox}
    
    \vspace{2pt}

    \begin{tcolorbox}[
        enhanced,
        colback=orange!1!white,             
        colframe=orange!50!black,
        boxrule=0.5pt,
        arc=2pt,
        top=10pt, bottom=4pt, left=6pt, right=6pt, 
        fontupper=\scriptsize\ttfamily\linespread{1.1}\selectfont\raggedright, 
        title={\sffamily\bfseries User Prompt},
        attach boxed title to top left={xshift=10pt, yshift=-8pt},
        boxed title style={colback=orange!85!black, colframe=orange!85!black, size=small}
    ]

[Rule Information]
\par\vspace{3pt}
\textcolor{gray}{...(Omitted content)}
\par\vspace{3pt}

[Commit Information to Evaluate]
\par\vspace{3pt}
\textcolor{gray}{...(Omitted content)}
\par\vspace{3pt}

- Project Directory Tree (Snapshot after commit):
\par
{project\_tree\_structure}

\par\vspace{3pt}

- File Diffs (Code Changes in this Commit):
\par
{file\_diff\_contents}
\par
*(Note: Strictly judge the developer's behavior based on the '+' lines to avoid penalizing them for legacy code).*
\par\vspace{3pt}
Please analyze the data above and output the evaluation in the required JSON format.
    \end{tcolorbox}

    \end{tcolorbox}
    \caption{The prompt structure used for rule compliance assessment.}
    \Description{The figure presents the prompt structure used for rule compliance assessment. Similar to the previous design, the prompt consists of a system prompt and a user prompt. The system prompt defines the LLM as a rigorous empirical software engineering auditor and outlines a step-by-step evaluation procedure. First, the model classifies the rule as either quantifiable or boolean to determine the evaluation granularity. It then performs an applicability check to determine whether the given commit is relevant to the rule, preventing incorrect judgments on unrelated changes. Based on this, the model analyzes the project directory snapshot and code diffs associated with the commit, focusing strictly on added lines (``+'' lines) to evaluate developer behavior. The user prompt provides both the rule information and the commit data. Finally, the model outputs a structured JSON result indicating whether the commit complies with the rule, enabling scalable automated compliance analysis.}
    \label{fig:prompt_compliance}
\end{figure}

Finally, because recent studies show that using LLMs as judges in software engineering tasks achieves results comparable to human evaluation \cite{HeShZhTrSuXiDuLo2025, WaGuGaFaChXi2025}, we constructed an evaluation process based on the LLM-as-a-judge approach. We developed the prompt iteratively: we sampled a small set of records to draft an initial version, manually inspected the LLM outputs, identified errors, summarized their root causes, and refined the prompt accordingly. We repeated this cycle until three consecutive rounds of manual inspection revealed no errors or unreasonable judgments. Figure \ref{fig:prompt_compliance} presents the resulting abridged prompt (the full version is available in our replication package \cite{dataset}). To reduce hallucinations, the LLM evaluated each commit independently. As shown in the User Prompt of Figure \ref{fig:prompt_compliance}, we supplied the LLM with inputs including rule-specific details, target commit information, a snapshot of the project directory tree after the commit, and all file Diffs from that commit. We defined specific evaluation logic and output constraints: the LLM was required to first perform an applicability check to bypass commits irrelevant to the rule's target artifacts. During this step, if the LLM found no checkable compliance points in all preceding or all succeeding commits for a rule, we excluded that rule (removing 80 rules). Furthermore, the LLM produced structured JSON output that explicitly distinguishes between boolean rules, which verify a global state, and quantifiable rules, which evaluate each modified instance individually. We executed this evaluation pipeline on the 240 rules using Gemini 3 Flash Preview. Upon completion, we conducted a manual inspection of the 160 remaining generated results to ensure the validity of the automated evaluations.

\subsection{Survey Design}\label{survey_design}
\begin{figure*}[t]
\centering
\includegraphics[
    width=\textwidth,
]{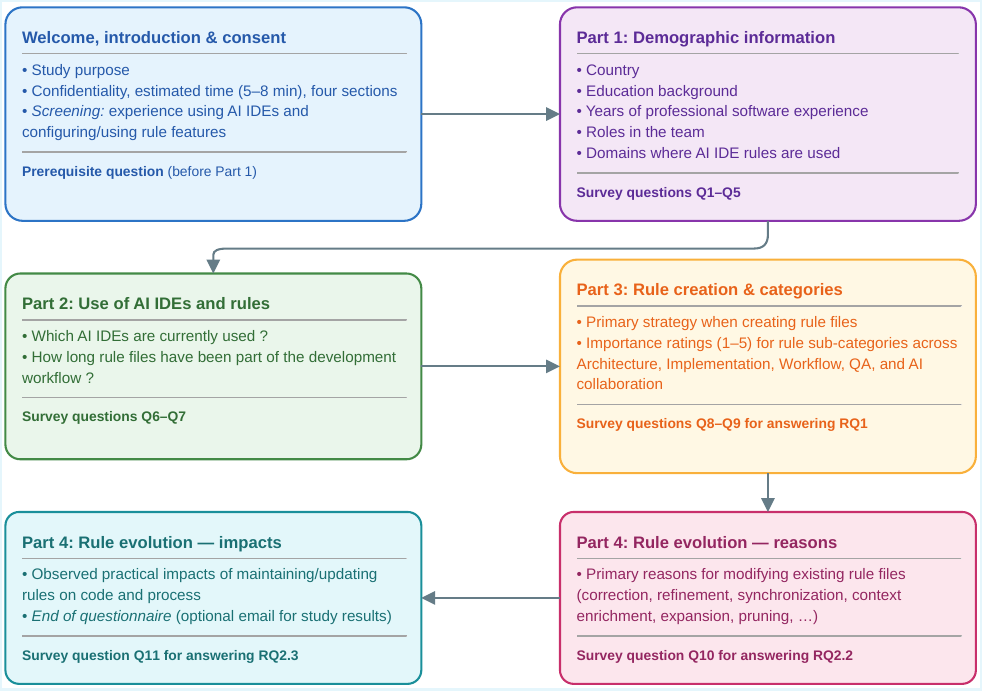}
\caption{Overview of the survey questionnaire.}
\Description{A flowchart illustrating the structure and flow of the survey questionnaire. The survey is divided into four sequential parts. Part 1 collects demographic information and general programming experience. This is followed by a screening question: if the participant has no experience using AI IDEs, the survey terminates. Eligible participants proceed to Part 2, which investigates their rule configuration practices, matching the taxonomy discussed in RQ1. Part 3 explores rule evolution and maintenance, asking participants about their motivations for updating or deleting rules, corresponding to RQ2. The final part, Part 4, collects open-ended feedback and allows participants to opt-in for follow-up interviews.}
\label{fig:survey_workflow}
\end{figure*}

The second phase of our study is a practitioner survey designed to investigate how developers configure and maintain rules in AI IDEs in practice. This survey aims to achieve methodological triangulation by complementing the repository mining findings from Section \ref{mining_study_design} with developers' subjective perceptions. The complete workflow of the survey is illustrated in Figure \ref{fig:survey_workflow}. To reach a broad and diverse audience, we administered the survey online. The questionnaire comprises single-choice questions, multiple-choice questions, and Likert-scale rating questions.

\subsubsection{Questionnaire Design}\label{questionnaire_design}
We designed the questionnaire to strictly align with our core objectives, ensuring that each Survey Question (SQ) maps directly to our Research Questions (RQs). The Welcome page of the survey outlines the background and purpose of the study. At the bottom of this page, we implemented a screening question (SQ0) to verify whether the respondent had practical experience using AI IDEs and their rule features. Only participants who answered affirmatively could proceed; otherwise, the survey terminated immediately. The main body of the questionnaire consists of four sections:
\begin{enumerate}
    \item \textit{Demographics:} Capturing basic background information, including location, education, professional experience, team role, and development domains.
    \item \textit{AI IDE and Rule Usage:} Identifying the specific AI IDEs used by participants and the duration of their experience with rule files.
    \item \textit{Rule Creation and Categories (mapping to RQ1):} Investigating strategies for rule creation and asking developers to rate the importance of various rule categories.
    \item \textit{Rule Evolution (mapping to RQ2):} Exploring the specific triggers for modifying rules (RQ2.2) and the practical impacts these changes have on their development processes (RQ2.3).
\end{enumerate}

The survey questions are outlined in Table \ref{tab:survey_questions} (the questionnaire is also available online\footnote{\url{https://forms.gle/sTTcK2cvt4UzEdSL6}}). Before finalization, the first and third authors collaboratively discussed the wording of each question to resolve any ambiguities. We then conducted two rounds of pilot surveys. In the first round, we sent the draft survey questionnaire to 20 contributors of GitHub projects containing rule files, received 2 valid responses, and refined the survey logic. In the second round, we sent the revised survey questionnaire to another 30 contributors, received 3 valid responses, and further optimized the formatting and phrasing to maximize readability.

\begin{table}[htbp]
\centering
\caption{Overview of the Survey Questionnaire}
\label{tab:survey_questions}
\scriptsize
\begin{tabularx}{\linewidth}{@{}llXl@{}}
\toprule
\textbf{ID} & \textbf{Section} & \textbf{Question} & \textbf{Type} \\ \midrule
SQ0 & Introduction & Do you have experience utilizing AI IDEs (e.g., Cursor, Windsurf) AND configuring their rule features (e.g., \texttt{.cursorrules})? & Single Choice \\ \addlinespace
SQ1 & Demographic & What is your primary location of work or study? & Single Choice \\
SQ2 & Demographic & What is your highest level of education? & Single Choice \\
SQ3 & Demographic & How many years of professional experience do you have in software development? & Single Choice \\
SQ4 & Demographic & What is your primary role in your team or organization? & Multiple Choice \\
SQ5 & Demographic & In which domains have you utilized AI IDEs and their rule features? & Multiple Choice \\
SQ6 & Demographic & Which AI IDEs do you currently use for development? & Multiple Choice \\
SQ7 & Demographic & How long have you been incorporating rule files into your development workflow? & Single Choice \\ \addlinespace
SQ8 & RQ1 & When creating rule files, what is your primary strategy? & Single Choice \\
SQ9 & RQ1 & Please rate the importance of 25 rule sub-categories across 5 primary categories. \newline\textit{(1 = completely unimportant/rarely used; 5 = crucial/essential)} & Likert Scale \\ \addlinespace
SQ10 & RQ2.2 & What are the primary triggers that prompt you to modify existing rule files? & Multiple Choice \\ \addlinespace
SQ11 & RQ2.3 & What practical impacts have maintaining and updating these rules had on your project code or development process? & Multiple Choice \\ \bottomrule
\end{tabularx}
\end{table}

\subsubsection{Data Collection and Filtering}\label{survey_data_collection}
To invite developers with hands-on experience modifying rule files in AI IDEs, we queried the GitHub API to identify repositories containing commits that touched rule files or directories between June 2025 and January 2026. From these commits, we collected the publicly available email addresses configured by the authors in their Git metadata. We carefully removed duplicates across different repositories and applied regular expressions to filter out invalid or auto-generated addresses (e.g., \texttt{users.noreply.github.com}, \texttt{localhost}). Following this process across the 5 selected AI IDEs, we identified a total of 2,302 unique developer emails. 

To ensure ethical compliance and respect for developer privacy, all data collection adhered to GitHub's Terms of Service regarding public information\footnote{\url{https://docs.github.com/en/site-policy/github-terms/github-terms-of-service}}. The collected emails were used exclusively for sending survey invitations. Participation in the survey was entirely voluntary and anonymous, with no personally identifiable information (PII) linked to the survey responses. Furthermore, the invitation emails included a clear opt-out mechanism, and the mailing list was permanently destroyed upon the conclusion of the data collection phase.

To maximize the response rate, we provided a localized version of the survey. Based on email suffixes, author names, and repository description languages, we identified 353 Chinese developers and sent them a translated Chinese version of the questionnaire. Out of the 2,302 emails dispatched, 162 bounced back, resulting in 2,140 effectively delivered invitations. Participation was entirely voluntary, and no financial compensation was provided. By the cutoff date of 20 April 2026, we received 117 initial responses (57 in English and 60 in Chinese).

To ensure data quality, we conducted a rigorous filtering process on the collected responses:
\begin{enumerate}
    \item \textit{Screening Failure:} 12 respondents (2 in English, 10 in Chinese) who answered ``No'' to SQ0 (indicating a lack of experience with AI IDE rules) were immediately excluded.
    \item \textit{Careless Responses:} We identified and excluded 6 careless responses (2 in English, 4 in Chinese) that exhibited ``straight-lining'' behavior. These participants gave the exact same score for all 25 rule categories in SQ9 AND blindly selected every available option in both SQ10 and SQ11, rendering their feedback meaningless for discussion.
\end{enumerate}

After applying these exclusion criteria, we obtained a final dataset of 99 valid responses (53 in English and 46 in Chinese) for our subsequent analysis.

\subsection{Mining Study Overview}
To comprehensively understand the ecosystem of projects driven by AI IDEs, we conducted a statistical analysis of the fundamental characteristics of the 83 collected open-source repositories developed using AI IDEs. This overview encompasses the primary programming languages, application domains, project scales, and rule configuration densities.

\begin{figure*}[t]
\centering
\includegraphics[
    width=\textwidth,
    clip
]{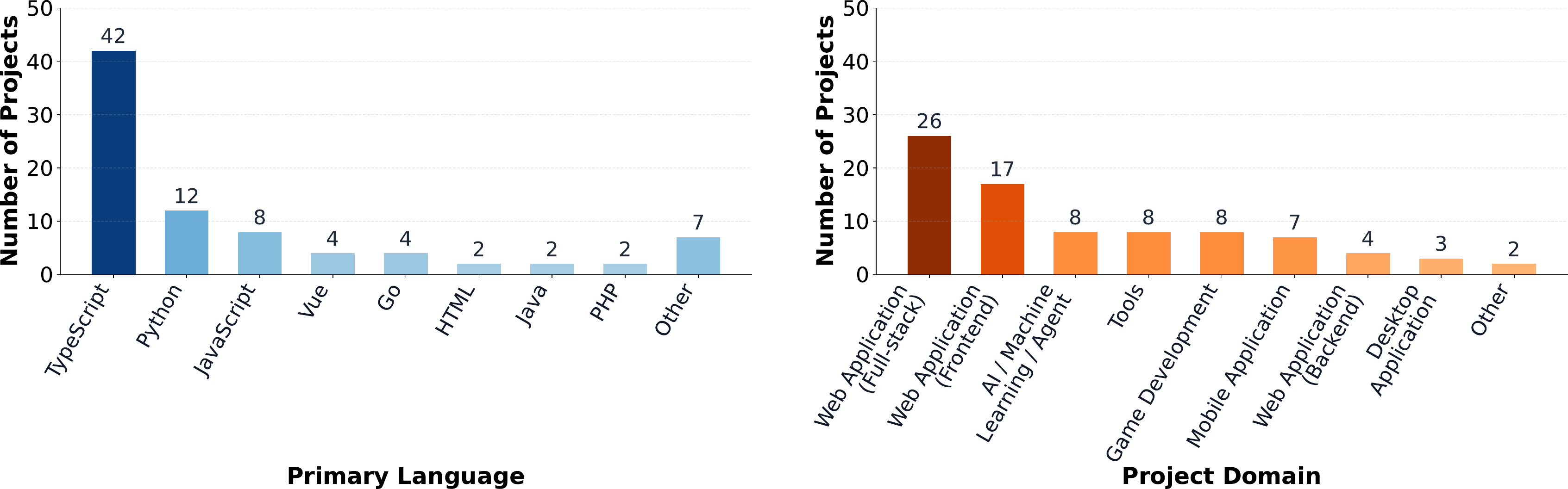}
\caption{Distribution of primary languages and project domains in the selected projects.}
\Description{The figure has two side-by-side vertical bar charts with a shared y-axis (``Number of Projects'', 0–50 in steps of 10). Left (Primary language): counts use the first language listed per repository; TypeScript is most common (42), then Python (12) and JavaScript (8); Vue and Go each appear 4 times; HTML, Java, and PHP each 2; the remaining 7 low-frequency languages are grouped as Other. Right (Project domain): Web Application (Full-stack) leads (26), followed by Web Application (Frontend) (17); AI / Machine Learning / Agent, Tools (merged developer-utility and media/creative domains), and Game Development each have 8; Mobile Application has 7; Web Application (Backend) has 4; Desktop Application has 3; 2 repositories fall in Other. Bars use blue (language) and orange (domain) gradients with counts printed on top.}
\label{fig:language_and_domain}
\end{figure*}

\textit{Languages and Domains.} As shown in the left panel of Figure \ref{fig:language_and_domain}, because some projects use multiple programming languages, we classified each repository according to its predominant PL. TypeScript dominates the dataset, accounting for 50.6\% (42 projects), followed by Python with 12 projects (14.4\%) and JavaScript with 8 projects (9.6\%). The ``Other'' category comprises a diverse array of languages, including Elixir, Rust, Dart, C\#, Kotlin, and Swift. Regarding application domains (in the right panel of Figure \ref{fig:language_and_domain}), Web Application development constitutes the overwhelming majority. Specifically, Full-Stack (26, 31.3\%), Frontend (17, 20.4\%), and Backend (4, 4.8\%) projects cumulatively account for 56.6\% of the collected projects. This suggests that Web engineering, with its strong emphasis on rapid iteration, is currently the most common adoption scenario for AI IDEs.

\begin{figure*}[t]
\centering
\includegraphics[
    width=\textwidth,
    clip
]{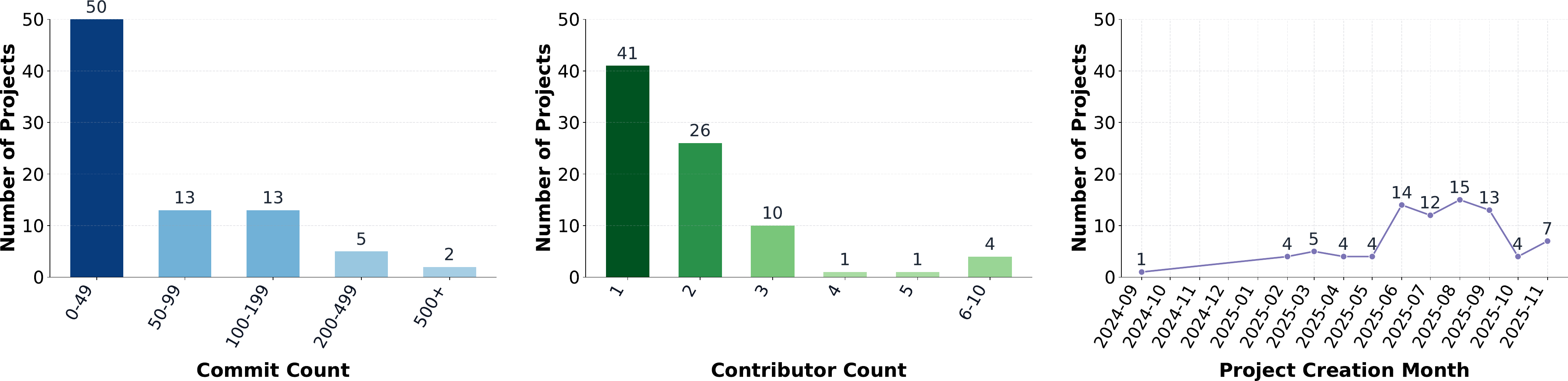}
\caption{Distribution of commit count, number of contributors, and project creation month in the selected project.}
\Description{Three panels share one y-axis (“Number of Projects”). Left (Commit count): repositories are binned by total commit count; most (50) have 0–49 commits, 13 each in 50–99 and 100–199, 5 in 200–499, and 2 with 500+ commits. Middle (Contributor count): 41 repos have a single contributor, 26 have two, 10 have three; one repo each has four and five contributors, and 4 have six to ten. Right (Project creation month): a line with markers counts first commits by calendar month (from ISO timestamps); activity is low before 2025, then rises through 2025 with the largest monthly counts in June (14), August (15), July (12), and September (13), and smaller counts in other months (down to 1 in 2024-09). Numeric labels sit above each point.}
\label{fig:git_stats}
\end{figure*}

\textit{Project Scale and Creation Time.} Figure \ref{fig:git_stats} presents the distribution of commit volumes, contributor counts, and project creation dates. Notably, projects developed entirely using AI IDEs are currently predominantly small-to-medium in scale. Over half of the projects (50, 60.2\%) have fewer than 50 historical commits. In terms of team size, solo developers or micro-teams of 1 to 3 people account for the vast majority (1 person: 49.4\%, 2 people: 31.3\%, 3 people: 12.0\%). This characteristic provides a profound reflection of the current landscape of projects developed by AI IDEs: they are primarily driven by independent developers or extremely small teams leveraging AI for agile development. Furthermore, based on the timestamp of the first commit, the rightmost chart shows that many of these projects were created between June 2025 and September 2025. This timeline closely aligns with the recent rapid updates and widespread adoption of mainstream AI IDE tools.

\begin{figure*}[t]
\centering
\includegraphics[
    width=\textwidth,
    clip
]{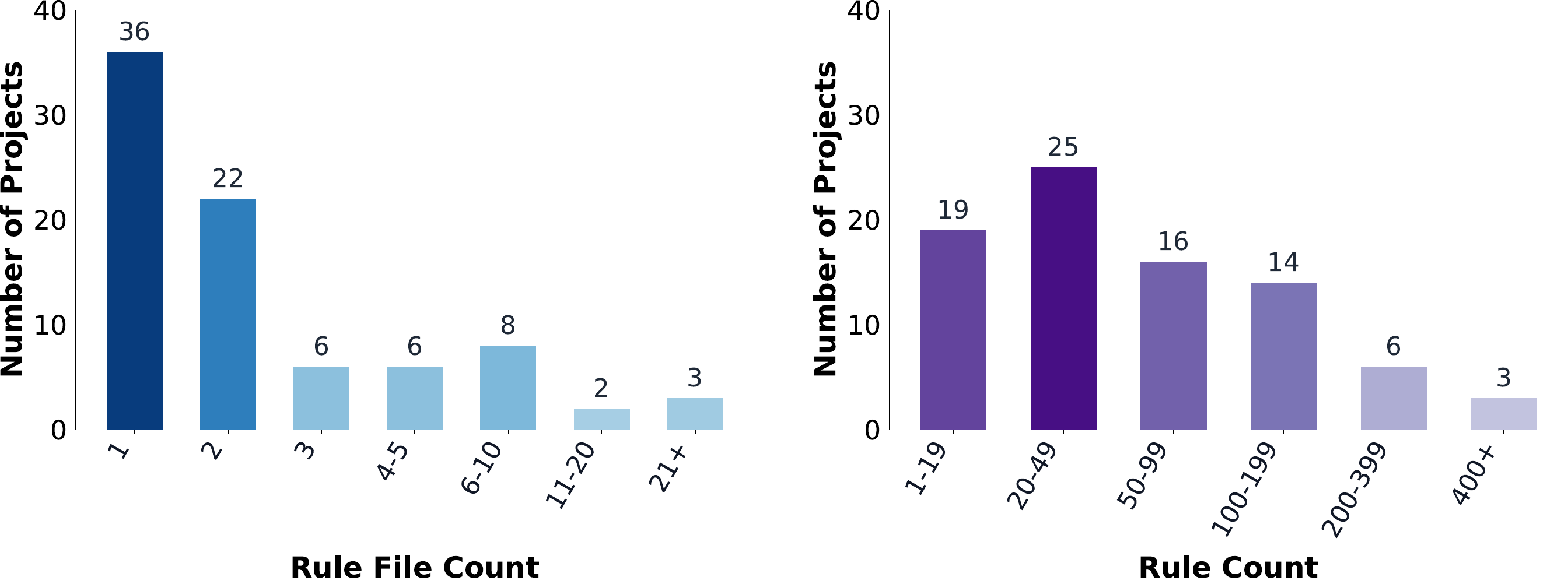}
\caption{Overview of the number of rule files and rule entries in each project.}
\Description{Two side-by-side vertical bar charts share the y-axis (``Number of Projects''). Left (Rule file count): 36 repositories have exactly one rule JSON file, 22 have two, and 6 have three; 6 fall in 4–5 files, 8 in 6–10, 2 in 11–20, and 3 have 21+ files. Right (Rule count, total objects across files): the modal band is 20–49 objects (25 repos), followed by 1–19 (19), 50–99 (16), and 100–199 (14); 6 repos have 200–399 objects and 3 have 400+. Bars use blue and purple gradients, respectively, with counts above each bar.}
\label{fig:rule_file_and_rules}
\end{figure*}

\textit{Rule Distribution.} Figure \ref{fig:rule_file_and_rules} reveals the distribution of rule files and individual segmented rules across the projects. The left chart indicates that most projects favor a centralized configuration pattern: projects with exactly 1 or 2 rule files account for 43.4\% (36) and 26.5\% (22), respectively. Repositories maintaining an extensive number of rule files (over 20) are extremely rare, comprising only 3.6\% (3 projects). Following our semantic segmentation (see Section \ref{mining_rule_extraction_taxonomy}), the right chart displays the total number of individual rules per project. Projects containing 20\textasciitilde49 rules are the most common (25, 30.1\%), followed by those with 1\textasciitilde19 rules (19, 22.9\%), 50\textasciitilde99 rules (16, 19.3\%), and 100\textasciitilde199 rules (14, 16.9\%). Similarly, projects maintaining a massive context of over 400 rules are scarce (3, 3.6\%). These distributions suggest that developers generally prefer to author mid-scale (dozens to a few hundred), high-density core rule sets for their AI assistants, rather than heavily fragmenting rule configurations across numerous files.

\subsection{Survey Demographics}
\begin{figure*}[t]
\centering
\includegraphics[
    width=\textwidth,
    trim=2cm 0cm 2cm 1cm,
    clip
]{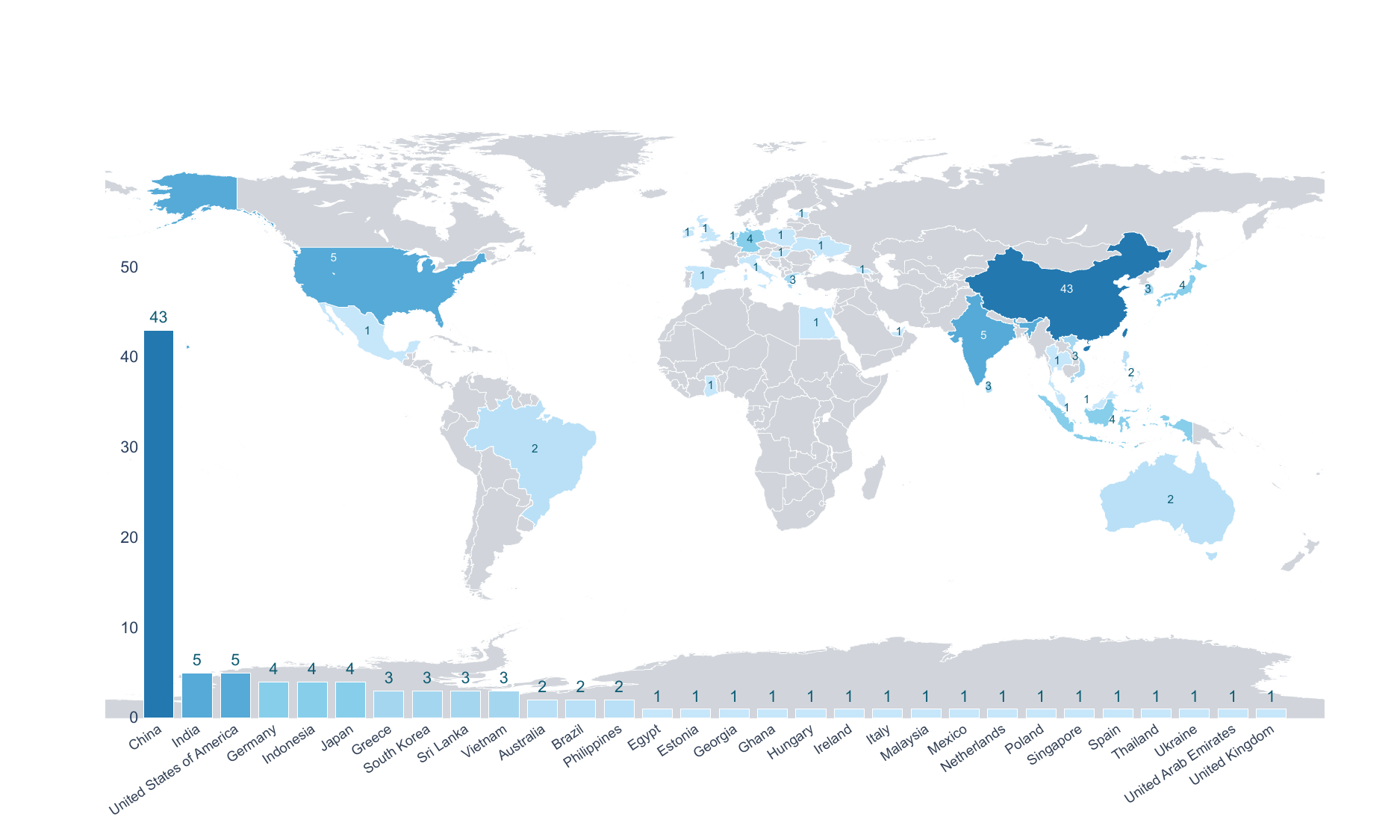}
\caption{Overview of countries of survey participants.}
\Description{The figure shows data from respondents in 31 countries/regions. The number of respondents is as follows: China 43, India 5, USA 5, Germany 4, Indonesia 4, Japan 4, Greece 3, South Korea 3, Sri Lanka 3, Vietnam 3, Brazil 2, Philippines 2, Australia 2; and 1 each from Egypt, Estonia, Georgia, Ghana, Hungary, Ireland, Italy, Malaysia, Mexico, Netherlands, Poland, Spain, Thailand, Ukraine, UAE, UK, and Singapore.}
\label{fig:participants_world_map}
\end{figure*}

Our survey captured a diverse global sample of software practitioners. The 99 valid responses came from 30 different countries and regions. The top represented countries include China (43, 43.4\%), India (5, 5.1\%), and the United States (5, 5.1\%), followed by Germany (4, 4.0\%), Indonesia (4, 4.0\%), and Japan (4, 4.0\%). A complete geographical distribution of the participants is presented in Figure \ref{fig:participants_world_map}.

\begin{figure*}[t]
\centering
\includegraphics[
    width=\textwidth,
]{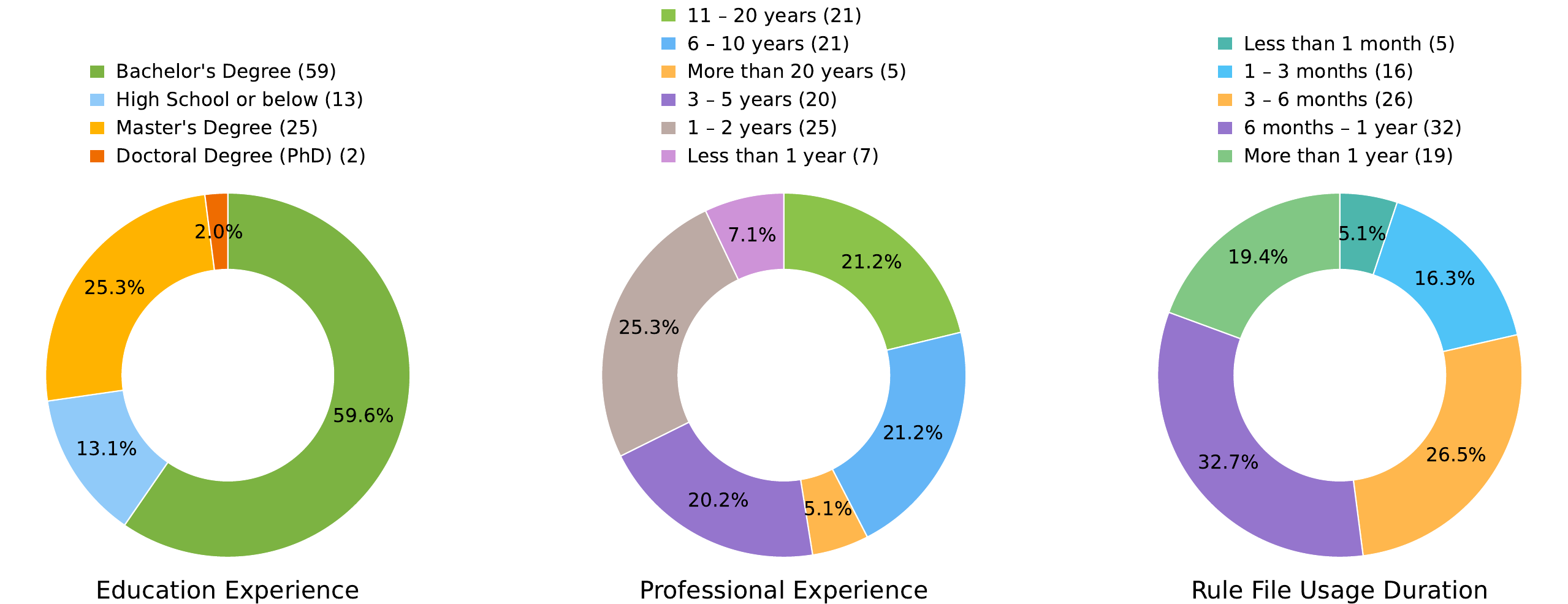}
\caption{Overview of education, professional experience and rule file usage duration of survey participants.}
\Description{This figure includes statistics from two groups of respondents (total number 99). The left side shows the educational background distribution: Bachelor's degree 59 (59.6\%), Master's degree 25 (25.3\%), high school diploma or below 13 (13.1\%), and PhD 2 (2.0\%). The right side shows the software development experience distribution: 1-2 years 25 (25.3\%), 6-10 years 21 (21.2\%), 11-20 years 21 (21.2\%), 3-5 years 20 (20.2\%), less than 1 year 7 (7.1\%), and over 20 years 5 (5.1\%). Overall, Bachelor's degree holders accounted for the largest proportion, and development experience was mainly concentrated in the 1-2 year, 3-5 year, 6-10 year, and 11-20 year range.}
\label{fig:education_professional_duration}
\end{figure*}

Figure \ref{fig:education_professional_duration} illustrates the participants' educational backgrounds, professional software development experience, and the duration of their experience in configuring AI IDE rules. Regarding education, the vast majority of respondents (59, 59.6\%) hold a Bachelor's degree, while only 2 respondents hold a PhD. In terms of professional experience, the distribution is relatively even and leans toward mid-to-senior levels: those with 1--2 years, 3--5 years, 6--10 years, and 11--20 years of experience account for 25.3\% (25), 20.2\% (20), 21.2\% (21), and 21.2\% (21) of the sample, respectively. Beginners with less than 1 year (7) and veterans with over 20 years of experience (5) constitute a minority. Concerning the duration of rule file usage, the largest segment of respondents has been incorporating rules into their workflows for 6 months to one year (32), followed by 3 to 6 months (26). Notably, 19 respondents are early adopters with over one year of experience, whereas only 5 respondents have used rule features for less than a month.

\begin{figure*}[t]
\centering
\includegraphics[
    width=\textwidth,
]{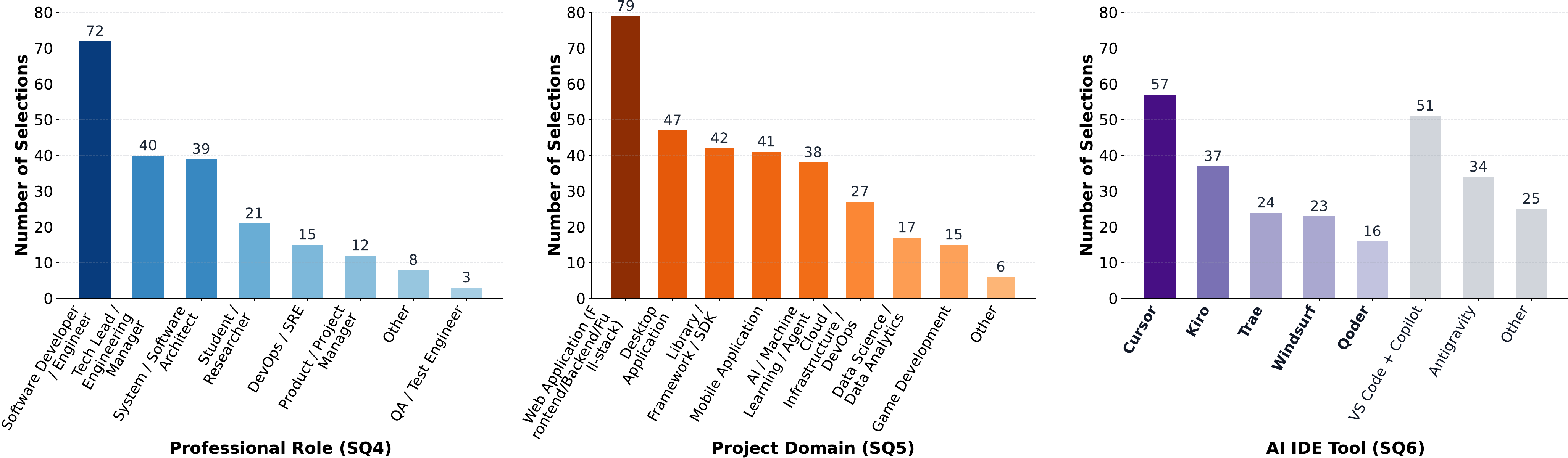}
\caption{Overview of roles in the team, project domains developed using the AI IDE, and AI IDE usage of survey participants.}
\Description{This figure contains three parallel bar charts, showing the results of SQ6 of the questionnaire. The vertical axis represents the number of times each option was selected (multiple-choice questions, so the total number of selections for each category may exceed the number of respondents). The left bar chart for SQ4 (team role) shows: Software Developer/Engineer (72 selections), followed by Tech Lead/Engineering Manager (40 selections), System/Software Architect (39 selections); Student/Researcher (21 selections), DevOps/SRE (15 selections), Product/Project Manager (12 selections), Other (8 selections), and QA/Test Engineer (3 selections). The middle bar chart for SQ5 (application area) shows: Web Application (79 selections), followed by Desktop Application (47 selections), Library/Framework/SDK (42 selections), Mobile Application (41 selections), AI/Machine Learning/Agent (38 selections); Cloud/Infrastructure/DevOps (27 selections), Data Science/Data Analytics (17 selections), Game Development (15 selections), and Other (6 selections). The right-hand figure SQ6 (IDEs used) shows: Cursor 57 times, VS Code + Copilot 51 times, Kiro 37 times, Antigravity 34 times, Trae 24 times, Windsurf 23 times, Qoder 16 times, and Other 25 times; among them, the five IDEs that are the focus of the research (Cursor, Windsurf, Trae, Qoder, and Kiro) are highlighted in the figure.}
\label{fig:role_domain_ide}
\end{figure*}

Figure \ref{fig:role_domain_ide} details the participants' team roles, their development domains utilizing AI IDEs, and the specific tools they have adopted (note that these three questions allow multiple selections). For team roles, a dominant 72.7\% (72) of the participants act as \textit{Software Developers/Engineers}. Additionally, a substantial portion hold leadership or architectural positions, comprising 40 \textit{Tech Leads/Engineering Managers} and 39 \textit{System/Software Architects}. \textit{Students/Researchers} account for 21 respondents. Conversely, only 3 respondents are \textit{QA/Test Engineers}, which may imply that testing professionals currently engage with generative AI coding tools less extensively than core developers. The ``Other'' category (8 respondents) further broadens our demographic, including roles such as \textit{Creative Director}, \textit{VFX Supervisor}, \textit{Technical Solo Developer}, \textit{UX Designer}, and \textit{Founder}. 

In terms of project domains, \textit{Web Application} development is overwhelmingly prevalent (79), followed by \textit{Desktop Applications} (47), \textit{Libraries/Frameworks/SDKs} (42), \textit{Mobile Applications} (41), and \textit{AI/Machine Learning/Agent} (38). \textit{Cloud/DevOps/Infrastructure} (27), \textit{Data Science/Data Analytics} (17), and \textit{Game Development} (15) were less frequently selected. The ``Other'' field includes specialized areas such as drone communication and IoT, robotics simulation, and technical prototyping. Regarding tool preferences, \textit{Cursor} emerges as the most widely adopted AI IDE among our respondents (57), closely followed by the traditional \textit{VS Code + Copilot} setup (51). The other four selected AI IDEs in our study---\textit{Kiro}, \textit{Trae}, \textit{Windsurf}, and \textit{Qoder}---were utilized by 37, 24, 23, and 16 participants, respectively. Furthermore, 34 respondents reported using \textit{Antigravity}, while tools like \textit{Claude Code} (10), \textit{Codex} (7), and \textit{OpenCode} (4) were specified under the ``Other'' option (the detailed responses for all ``Other'' options across these figures are available in our replication package \cite{dataset}).

\section{Study Results}\label{sec_result}

\subsection{RQ1: What are the categories of rules in OSS projects developed by AI IDEs?}\label{sec_rq1_result}
To answer RQ1, we present our findings in three parts. First, we introduce the hierarchical rule taxonomy, comprising 5 primary and 25 secondary categories derived from the mining data with the open coding process (Section \ref{sec_rule_taxonomy_overview}). Next, we characterize this taxonomy by comparing the prevalence of rules in the repositories with the importance ratings from developers (Section \ref{sec:rq1_quantitative_analysis}). Finally, we report on the strategies developers employ to create these rule files in practice (Section \ref{sec:rq1_creation_strategies}).

\subsubsection{Overview of the Rule Taxonomy}\label{sec_rule_taxonomy_overview}
Through an open coding process on the collected rules, we constructed a hierarchical taxonomy comprising 5 primary categories and 25 secondary categories. In this section, we describe each category and illustrate its application using a representative example extracted from the repositories.

\vspace{1ex}
\noindent \textbf{1. Architecture \& Design.} 
This primary category encompasses constraints regarding the high-level structure, technology selection, and macro-design philosophies of the project.
\begin{itemize}
    \item \textbf{Technology Stack Selections.} Defines the core languages, frameworks, libraries, databases, and infrastructure for the project. \\ Example: ``\textit{The project uses the following technologies: TypeScript, Next.js (App Router), Supabase for Database \& Auth, Tailwind CSS with shadcn/ui components...}''.
    \item \textbf{Design Principles \& Patterns.} Specifies the software engineering principles (e.g., SOLID) and common design patterns adopted in the system. \\ Example: ``\textit{Apply SOLID principles, Clean Architecture, and Domain-Driven Design (DDD)}''.
    \item \textbf{System Architecture.} Describes the high-level structure and organizational relationship of the system. \\ Example: ``\textit{System architecture is Gateway $\rightarrow$ Kiosk $\rightarrow$ ModbusController $\rightarrow$ Hardware}''.
    \item \textbf{Design References \& Constraints.} Specifies external documentation, best practices, or specific constraints that the design must adhere to. \\ Example: ``\textit{Reference official AWS documentation for implementation patterns}''.
\end{itemize}

\vspace{1ex}
\noindent \textbf{2. Code Implementation.} 
This category focuses on concrete coding practices, guiding the AI IDE on how to write code at the file and function levels.
\begin{itemize}
    \item \textbf{Framework Usage.} Best practices and key considerations when utilizing specific frameworks. \\ Example: ``\textit{Always use the latest Vue 3 Composition API syntax}''.
    \item \textbf{Code Style Conventions.} Conventions for naming identifiers (e.g., variables, functions, classes) and overall code formatting. \\ Example: ``\textit{Use UPPER\_SNAKE\_CASE for constants}''.
    \item \textbf{Performance Optimization.} Guidelines aimed at improving code execution efficiency. \\ Example: ``\textit{Use prepared statements and indexes for query optimization}''.
    \item \textbf{Language Features.} Best practices and restrictions regarding specific programming language syntax and semantics. \\ Example: ``\textit{Channel Closing: Only close channels from the sender}''.
    \item \textbf{Error \& Exception Handling.} Unified mechanisms for error reporting, handling, and recovery. \\ Example: ``\textit{Implement a global error handling mechanism using an ErrorBoundary component that integrates with a reporting service like Sentry}''.
    \item \textbf{Business Logic.} Defines entity structures, business rules (e.g., validation logic), and specific functional processes within a specific domain to guide the AI IDE in accurately translating requirements into code. \\ Example: ``\textit{Registration with Email Verification: Users must verify email before account activation}''.
\end{itemize}

\vspace{1ex}
\noindent \textbf{3. Development Workflow \& Project Management.} 
This category addresses rules governing software lifecycle management, collaboration processes, and the organization of project assets.
\begin{itemize}
    \item \textbf{Workflow Conventions.} Standardized development processes (e.g., issue management, PR workflows, CI/CD pipelines). \\ Example: ``\textit{MANDATORY WORKFLOW BEFORE COMMITTING: \# 1. Run tests first make test \# 2. If tests pass, stage and commit git add . git commit -m `type(scope): description'}''.
    \item \textbf{Project Documentation.} Standards for writing and managing project documentation (e.g., API docs, user manuals). \\ Example: ``\textit{API Documentation must detail each endpoint, including URL, method, purpose, request/response formats, error codes, security, and rate limits}''.
    \item \textbf{Directory Structure.} Rules dictating directory hierarchy, file placement, and naming conventions for files/directories. \\ Example: ``\textit{One package per directory, main package in cmd/}''.
    \item \textbf{Environment Configuration.} Configuration rules for development environments, dependencies, and toolchains. \\ Example: ``\textit{Use environment variables for all API keys and configuration}''.
    \item \textbf{Version Control.} Conventions for code history management, versioning, and commit practices. \\ Example: ``\textit{Write descriptive commit messages following this format: First line: Concise summary (50 chars or less); Body: Detailed explanation if needed (wrap at 72 chars)}''.
    \item \textbf{Dependency Management.} Standards for importing libraries, resolving version dependencies, and package management. \\ Example: ``\textit{For external imports, always define a version. For example, npm:@express should be written as npm:express@4.18.2}''.
\end{itemize}

\vspace{1ex}
\noindent \textbf{4. Quality Assurance.} 
This category encompasses standards and practices designed to ensure software reliability, security, and maintainability.
\begin{itemize}
    \item \textbf{Testing Strategy.} Guidelines on how to design, write, and execute software tests. \\ Example: ``\textit{Write unit, integration, and e2e tests as appropriate}''.
    \item \textbf{Security Practices.} Defensive coding practices aimed at preventing security vulnerabilities. \\ Example: ``\textit{Use parameterized queries, never concatenate SQL}''.
    \item \textbf{Code Quality Standards.} Specific metrics and requirements for measuring and enforcing code quality. \\ Example: ``\textit{Code quality metric thresholds should be met: Code coverage >80\%, Cyclomatic complexity <10, Code duplication rate <5\%, Test execution time <5 minutes}''.
    \item \textbf{Logging Standards.} Requirements regarding the formats, severity levels, and locations for logging. \\ Example: ``\textit{Purposeful Logging: Every log entry must provide actionable value for operations or debugging}''.
    \item \textbf{Code Review.} Processes and standards for evaluating code, particularly when utilizing AI as a reviewer. \\ Example: ``\textit{Code Review Requirements: All API endpoints must have error handling}''.
\end{itemize}

\vspace{1ex}
\noindent \textbf{5. AI Collaboration Specifications.} 
This category acts as ``meta-controls'' to regulate the inherent behavior, interaction style, and tool usage of the Large Language Models in AI IDEs.
\begin{itemize}
    \item \textbf{AI Behavior \& Decision Strategies.} Controls the AI IDE's actions in specific situations, such as intent recognition, response logic, and mandatory safety checks (e.g., stopping upon encountering errors). \\ Example: ``\textit{Always verify information before presenting it. Do not make assumptions or speculate without clear evidence}''.
    \item \textbf{AI Output Content Guidelines.} Standardizes the AI IDE's response format during interactions, including language requirements, tone, depth of explanation, and structural layout. \\ Example: ``\textit{Minimal Output: Answer directly, avoid unnecessary preambles/postambles}''.
    \item \textbf{AI Context Management.} Defines the AI's persona and guides how it should maintain and utilize the project context. \\ Example: ``\textit{To maximize the effectiveness of this brief, reference specific Markdown files or sections based on the code you need the AI to generate}''.
    \item \textbf{AI Tool Usage.} Specifies the tools available to the AI IDE and imposes constraints on their invocation. \\ Example: ``\textit{You HAVE TO use generateTheme tool to generate the theme, do NOT just output XML type text for tool-call, that is not allowed}''.
\end{itemize}

\subsubsection{Distribution and Importance Analysis of Rule Categories}\label{sec:rq1_quantitative_analysis}
Table \ref{tab:rule_taxonomy} presents the quantitative distribution of the collected rules alongside the importance ratings provided by survey participants.

\begin{table*}[tbh] 
    \centering
    \caption{Taxonomy of AI IDE Rules and Elicited Importance from Survey}
    \label{tab:rule_taxonomy}    
    \includegraphics[width=\textwidth]{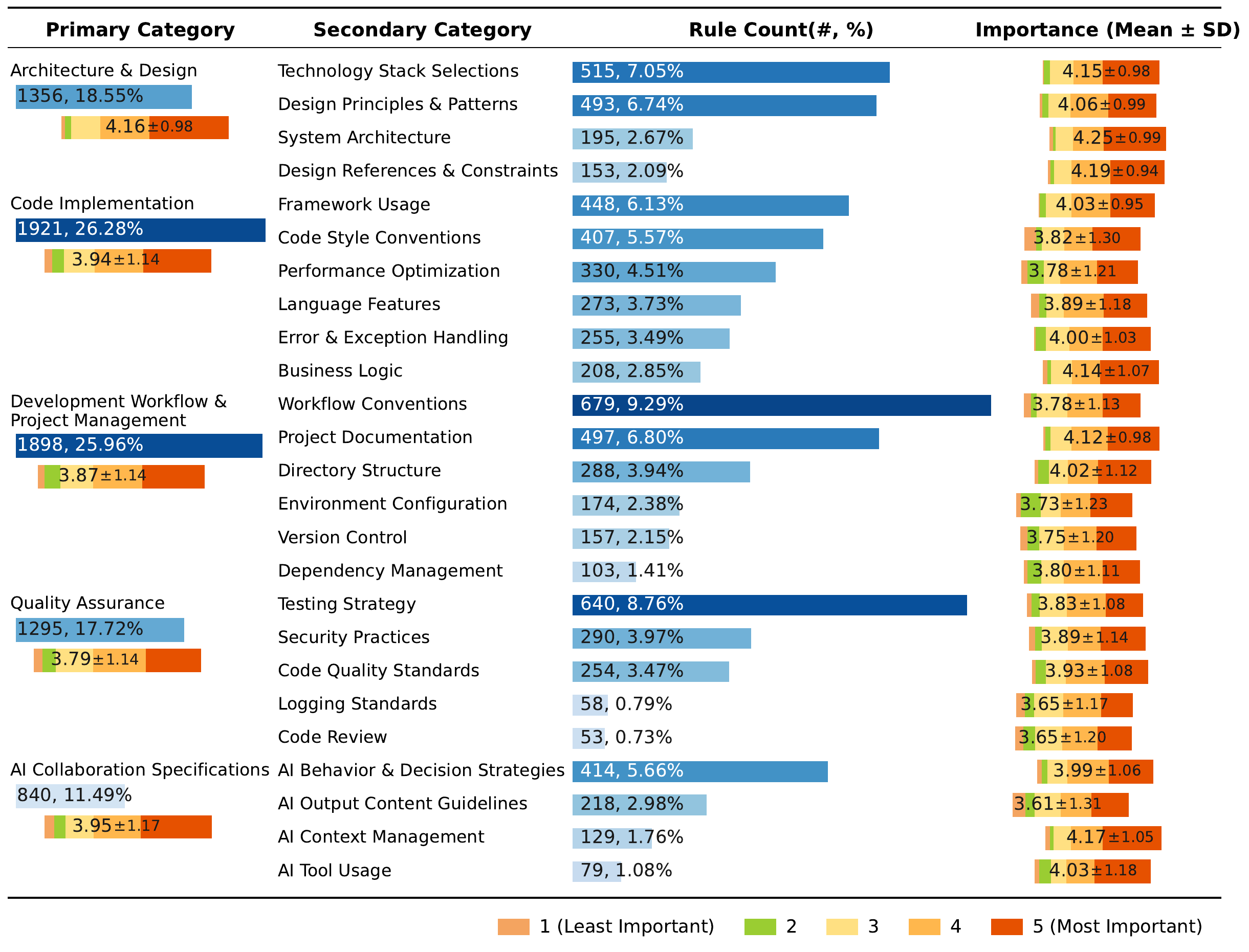}
\end{table*}

\textbf{Concentration Effect in Mining Repositories.} 
The distribution of rules across categories is uneven. \textit{Code Implementation} and \textit{Development Workflow \& Project Management} contain the highest numbers of rules, with 1,921 (26.28\%) and 1,898 (25.96\%) rules, respectively. In contrast, \textit{AI Collaboration Specifications} comprises the fewest rules (840, 11.49\%). Among the 25 secondary categories, \textit{Workflow Conventions} is the largest individual category with 679 rules (9.29\%), followed by \textit{Testing Strategy} with 640 rules (8.76\%). The distribution is also concentrated within specific primary categories. For instance, within \textit{Architecture \& Design}, the rules are concentrated in the top two subcategories: \textit{Technology Stack Selections} (515, 7.05\%) and \textit{Design Principles \& Patterns} (493, 6.74\%). A similar pattern exists in \textit{Quality Assurance}, which is primarily represented by \textit{Testing Strategy} (640, 9.76\%). Conversely, the smallest subcategories across the taxonomy include \textit{AI Tool Usage} (1.08\%), \textit{Logging Standards} (0.79\%), and \textit{Code Review} (0.73\%).

\textbf{Consensus and Divergence in Survey Perspectives.} 
The survey results show that developers prioritize structural and high-level constraints. \textit{Architecture \& Design} receives the highest overall importance rating (4.16), whereas \textit{Quality Assurance}---despite its high rule count---is rated the lowest (3.79). At the subcategory level, \textit{System Architecture} (4.25), \textit{Design References \& Constraints} (4.19), and \textit{AI Context Management} (4.17) receive the highest scores. Meanwhile, \textit{AI Output Content Guidelines} (3.61) and \textit{Logging Standards} (3.65) receive the lowest ratings. Furthermore, a Spearman's rank correlation test \cite{KuRe2023} between the mean importance scores and their standard deviations across the 25 subcategories reveals a statistically significant negative correlation ($\rho \approx -0.82, p < 0.001$). This indicates a higher consensus (i.e., lower variance) among developers regarding the most important rules, though this result may be partially influenced by the ceiling effect of the 1\textasciitilde 5 Likert scale, where average scores approaching the maximum limit inherently exhibit constrained variance.

\begin{figure*}[t]
\centering
\includegraphics[
    width=\textwidth,
]{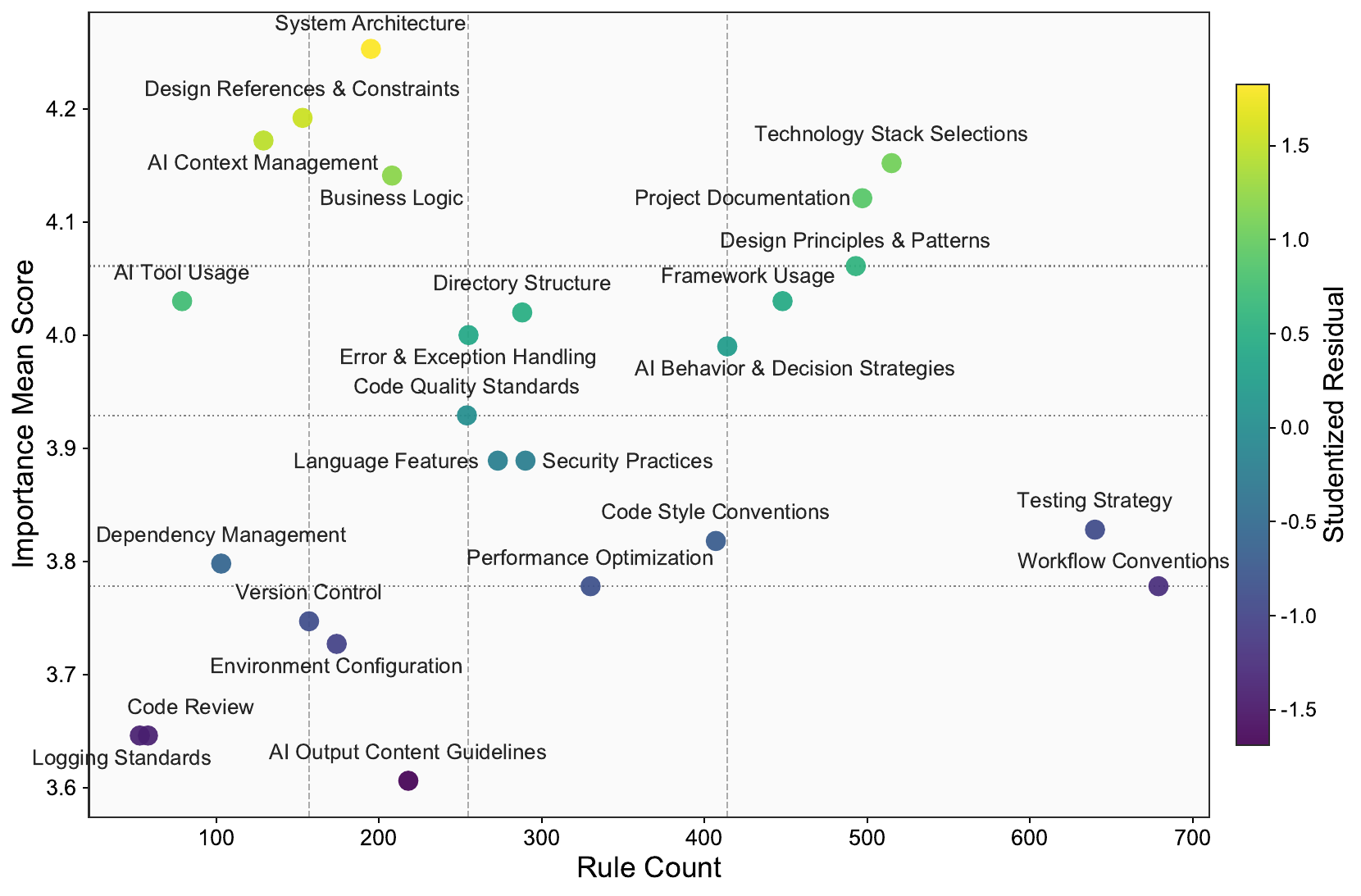}
\caption{Rule count and importance mean score across secondary categories.}
\Description{This is a two-dimensional scatter plot. The horizontal axis represents the ``Rule Count'' (the number of rules for each secondary category), and the vertical axis represents the ``Importance Mean Score'' (the average importance score of the secondary category in the questionnaire, Likert 1–5). The plot contains 25 solid dots, each corresponding to a secondary category name, with the text label directly above the dot. The color of the dots changes from dark purple to yellow-green along the color bar, representing the studentized residual obtained after performing a linear regression of the importance distribution on the number of rules using ordinary least squares: relative to the same fitted line, higher mean scores are more yellow-green, and lower mean scores are more purple; the color bar title is ``tudentized Residual''. The graph has no background grid; three vertical dashed lines represent the 25th, 50th, and 75th percentiles of the number of rules in this batch of 25 subcategories; three horizontal dotted lines represent the even distribution of importance among the 25th, 50th, and 75th percentiles of the importance in this batch of 25 subcategories. The coordinate axes are rectangular, with axes on the left, bottom, right, and top sides.}
\label{fig:scatter_rule_count_mean_score}
\end{figure*}

\textbf{Triangulation Analysis via Scatter and Quartiles.} 
To further explore the relationship between the prevalence of rules and developers' perceptions, we mapped the 25 subcategories onto a scatter plot (Figure \ref{fig:scatter_rule_count_mean_score}) with Rule Count on the X-axis and Importance Mean Score on the Y-axis. We performed an Ordinary Least Squares (OLS) regression residual analysis \cite{Bu2021} to identify categories deviating from the overall linear trend. The color of each point in Figure \ref{fig:scatter_rule_count_mean_score} maps to its studentized residual, which standardizes the distance of each data point from the predicted regression line, making it easier to visually identify extreme deviations. After applying the Bonferroni correction for multiple comparisons \cite{Ar2014}, no statistically significant outliers are identified. However, by overlaying quartile thresholds (the dashed lines in Figure \ref{fig:scatter_rule_count_mean_score}), we can analyze the distribution across quadrants. Using these quartile thresholds (where $Q_1$ and $Q_4$ denote the 1st and 4th statistical quartiles, respectively), we identified two main patterns of contrast:

\begin{itemize}
    \item \textbf{High-Prevalence, Low-Importance ($Q_4$ $\times$ $Q_1$):} \textit{Workflow Conventions} lie in the bottom-right quadrant. While it contains the highest number of rules ($Q_4$), its importance score falls into the lowest quartile ($Q_1$).
    \item \textbf{Low-Prevalence, High-Importance ($Q_1$ $\times$ $Q_4$):} Conversely, \textit{Design References \& Constraints} and \textit{AI Context Management} are located in the top-left quadrant. Despite containing relatively few rules ($Q_1$), they are rated within the highest importance quartile ($Q_4$).
\end{itemize}

\subsubsection{Rule Creation Strategies}\label{sec:rq1_creation_strategies}
To understand how developers create the rules in practice, we surveyed their primary strategies for creating rule files via a multiple-choice question in our survey (SQ8). The responses from the 99 participants are summarized in Table \ref{tab:rule_creation_strategies}.

\begin{table}[htbp]
    \centering
    \small
    \caption{Creation Strategies for AI IDE Rules ($N=99$, Multiple Choice)}
    \label{tab:rule_creation_strategies}
    \begin{tabular}{lcc}
        \toprule
        \textbf{Creation Strategy} & \textbf{\#} & \textbf{\%} \\
        \midrule
        Generate rules using AI and then refine them & 71 & 71.72\% \\
        Incremental addition (add rules only when AI makes mistakes) & 48 & 48.48\% \\
        Copy rules from other projects/templates and modify them & 38 & 38.38\% \\
        Write rules from scratch based on project needs & 35 & 35.35\% \\
        Others (automated scripts to combine/update rules, team rule sharing, etc.) & 2 & 2.02\% \\
        \bottomrule
    \end{tabular}
\end{table}

The survey data indicates that \textit{generating rules using AI and then refining them} is the most common strategy, selected by 71 respondents (71.72\%). The second most common approach is \textit{incremental addition}, where 48 respondents (48.48\%) reported adding new rules only when the AI IDE makes mistakes. Approximately one third of developers reported either \textit{copying and modifying rules from other projects or templates} (38, 38.38\%) or \textit{writing rules from scratch based on project needs} (35, 35.35\%).

Additionally, two respondents provided other strategies under the ``Other'' option. One respondent mentioned using community and organizational knowledge, stating that their rules are ``\textit{mainly derived from online sharing and company team research findings}''. The other response describes an automated practice for rule management: ``\textit{I created my own system using scripts to combine and update rules}''.

Overall, the survey results show that most developers rely on AI-assisted generation or an incremental approach when creating rules in AI IDEs. These findings suggest that rule creation is a dynamic process, motivating our investigation of rule evolution in RQ2.

\begin{tcolorbox}[
    enhanced,
    colback=gray!12!white,
    colframe=gray!50!black,
    arc=1.5mm,
    boxrule=1.4pt, 
    left=4pt, right=4pt, top=4pt, bottom=4pt
]
\textbf{Key Finding 1:} There is a notable difference between distribution of AI IDE rules in mining repositories and developers' perceived priorities. While developers highly value \textit{System Architecture} and \textit{AI Context Management} (High-Importance, Low-Prevalence), the analyzed projects are dominated by an abundance of rules concerning \textit{Workflow Conventions} and \textit{Code Style Conventions} (High-Prevalence, Low-Importance). Furthermore, rules rated with higher importance exhibit stronger consensus among developers (e.g., \textit{System Architecture} receives the highest mean score of 4.25 with a relatively low standard deviation of 0.99, whereas the lowest-rated \textit{AI Output Content Guidelines} shows higher variance with a standard deviation of 1.34). Additionally, rule creation is dynamic: most developers do not write rules from scratch, but instead rely on AI-assisted generation (71.72\%) or incremental additions based on generation errors (48.48\%).
\end{tcolorbox}

\subsection{RQ2: How do rules in OSS projects developed by AI IDEs evolve?}\label{sec_rq2_result}
Rules in AI IDEs are rarely static; instead, they change over time to adapt to evolving project requirements and mitigate AI limitations. To understand this evolution, we divide our analysis into three parts. First, we examine which categories of rules change and their specific change types (i.e., \textit{Added}, \textit{Modified} or \textit{Deleted}) (Section \ref{sec_rq2.1_result}). Building upon this, we investigate the motivations driving these changes, comparing reasons mined from commits with developers' reported intentions from the survey (Section \ref{sec_rq2.2_result}). Finally, we evaluate the impact of rule evolution by analyzing changes in the compliance of software artifacts following rule updates, alongside developers' perceived benefits (Section \ref{sec_rq2.3_result}).

\subsubsection{RQ2.1: What categories of rules undergo evolution?}\label{sec_rq2.1_result}

\begin{figure*}[t]
\centering
\includegraphics[
    width=\textwidth,
]{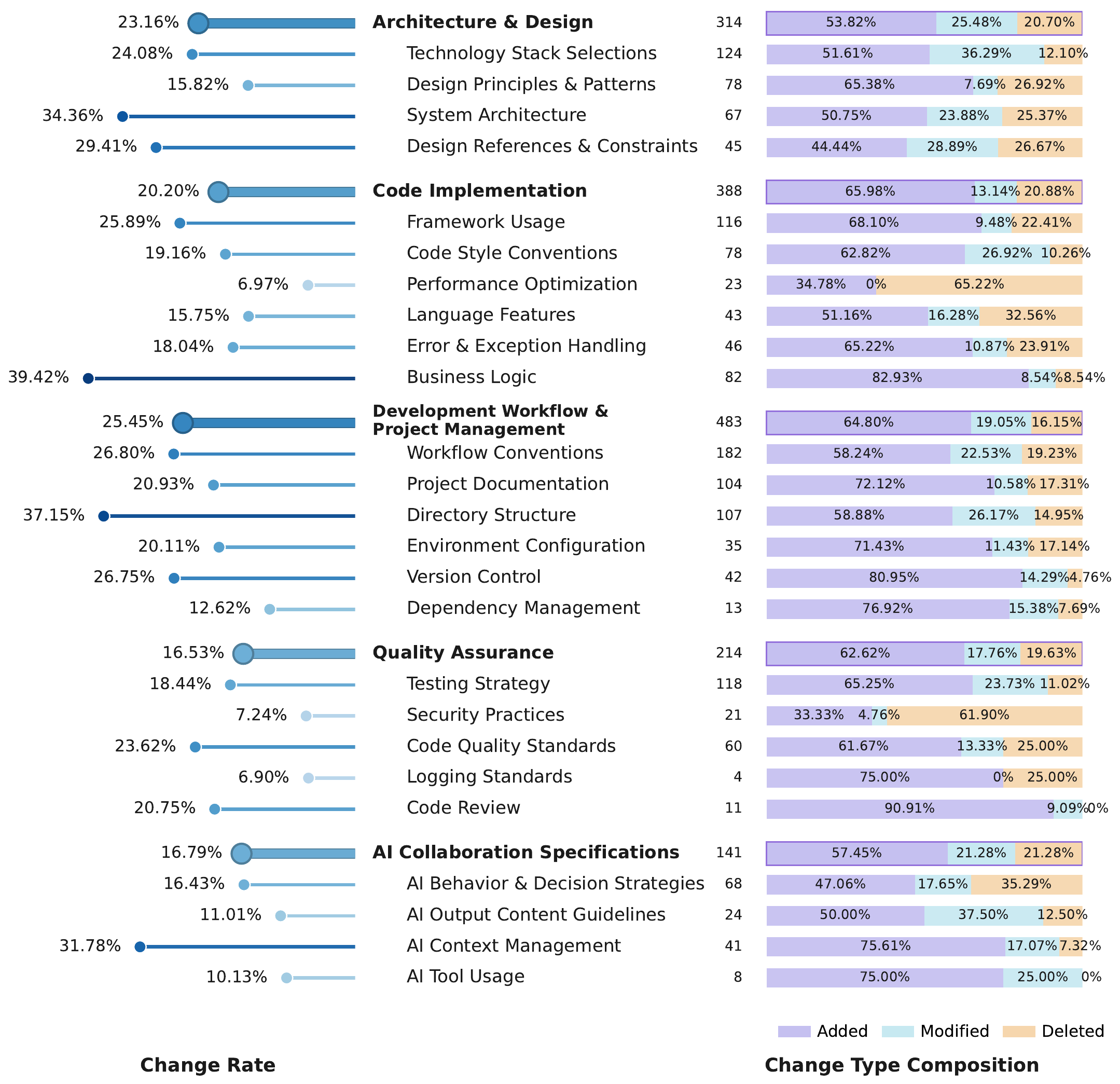}
\caption{Change rate and change type distribution of AI IDE rules.}
\Description{A horizontal dual-panel chart displaying the evolution metrics of 25 sub-categories of AI IDE rules across 5 primary categories. The left panel is a lollipop chart showing the Change Rate, where ``Business Logic'' and ``Directory Structure'' exhibit the highest mutation rates at over 37\%. The right panel is a 100\% stacked bar chart detailing the composition of changes (Added, Modified, Deleted). Across most categories, ``Added'' is the dominant change type, often exceeding 50\%, while ``Deleted'' accounts for the smallest proportion, except in specific categories like ``Performance Optimization'' where deletion is surprisingly prominent.}
\label{fig:rule_change_distribution}
\end{figure*}

Figure \ref{fig:rule_change_distribution} illustrates the distribution of rule changes across two dimensions. The central column lists the 5 primary and 25 secondary categories, annotated with the number of evolved rules ($N$). The left panel displays the change rate, representing the percentage of changed rules relative to the initial rule count for that category. The right panel uses a 100\% stacked bar chart to show the proportion of \textit{Added}, \textit{Modified}, and \textit{Deleted} operations.

\textbf{Variance in Change Rates.} 
At the primary category level, the data show that rules in several categories are frequently modified. \textit{Development Workflow \& Project Management} (Change Rate: 25.45\%, $N=483$), \textit{Architecture \& Design} (23.16\%, $N=314$), and \textit{Code Implementation} (20.20\%, $N=388$) are the three most frequently changed categories. In contrast, the change rates for \textit{Quality Assurance} (16.53\%) and \textit{AI Collaboration Specifications} (16.79\%) are lower.
Across the 25 secondary categories, the modification rates vary greatly. The four categories with the highest change rates are \textit{Business Logic} (39.42\%), \textit{Directory Structure} (37.15\%), \textit{System Architecture} (34.36\%), and \textit{AI Context Management} (31.78\%). Interestingly, cross-referencing this with RQ1 results reveals that although \textit{System Architecture} and \textit{Business Logic} do not have large initial rule counts, they change frequently during development, indicating that these rules require frequent human updates as the project progresses. Conversely, some categories rarely change once established, such as \textit{Logging Standards} (6.90\%), \textit{Performance Optimization} (6.97\%), and \textit{Security Practices} (7.24\%), all with change rates below 8\%.

\textbf{Distribution of Change Types.} 
As shown in the right panel of Figure \ref{fig:rule_change_distribution}, \textit{Added} is the most frequent operation across most categories (typically ranging from 50\% to 80\%). This demonstrates that rule evolution is primarily a process of adding new rules. However, \textit{Modified} and \textit{Deleted} operations show different distributions.
\begin{itemize}
    \item \textbf{Categories with High Modification Rates:} \textit{AI Output Content Guidelines} has the highest proportion of modifications at 37.50\%, followed by \textit{Technology Stack Selections} (36.29\%) and \textit{Design References \& Constraints} (28.89\%). This reflects areas where developers frequently refine rule texts to guide AI generation or align with specific libraries and frameworks versions. Notably, the modification rate is 0\% for \textit{Performance Optimization} and \textit{Logging Standards}.
    \item \textbf{Categories with High Deletion Rates:} While \textit{Deleted} is typically the least frequent operation in most categories, it is the dominant change type in \textit{Performance Optimization} (65.22\%) and \textit{Security Practices} (61.90\%), which are also the two categories with the lowest overall change rates. Additionally, \textit{AI Behavior \& Decision Strategies} shows a high deletion rate of 35.29\%.
\end{itemize}

\begin{tcolorbox}[
    enhanced,
    colback=gray!12!white,
    colframe=gray!50!black,
    arc=1.5mm,
    boxrule=1.4pt, 
    left=4pt, right=4pt, top=4pt, bottom=4pt
]
    \textbf{Key Finding 2:} AI IDE rules change frequently, particularly those related to \textit{Business Logic} and \textit{System Architecture}. Adding new rules (i.e., the \textit{Added} operation) is the dominant type of change across most rule categories, while edits to existing rules (i.e., the \textit{Modified} operation) are particularly common for \textit{AI Output Content Guidelines} and \textit{Technology Stack Selections}. In contrast, rules related to \textit{Security Practices} are rarely changed; however, when they are changed, they are typically \textit{Deleted}.
\end{tcolorbox}

\subsubsection{RQ2.2: What are the driving reasons for rule evolution?}\label{sec_rq2.2_result}
To identify the reasons driving rule evolution, we performed open coding on the 504 rule change records and classified them into six primary categories, as detailed below.
\begin{itemize}
    \item \textbf{Expansion:} Extending AI assistance scope to new project modules or adopting newly released features of the AI IDE.
    \item \textbf{Context Enrichment:} Providing implicit project background knowledge (e.g., hidden dependencies, architectural norms) to bridge the AI's context gap.
    \item \textbf{Synchronization:} Updating rules to maintain consistency with evolving codebases, such as refactored structures or upgraded tech stacks.
    \item \textbf{Refinement:} Optimizing the prompt phrasing or strategy to eliminate ambiguity and restrict overly broad AI behaviors, without changing the core intent.
    \item \textbf{Correction:} Enforcing strict negative constraints or corrections after observing AI hallucinations, buggy code generation, or undesirable behavioral patterns.
    \item \textbf{Pruning:} Deleting obsolete or redundant rules to streamline the context window and rely on improved capabilities of foundation models.
\end{itemize}

\begin{figure*}[t]
\centering
\includegraphics[
    width=\textwidth,
]{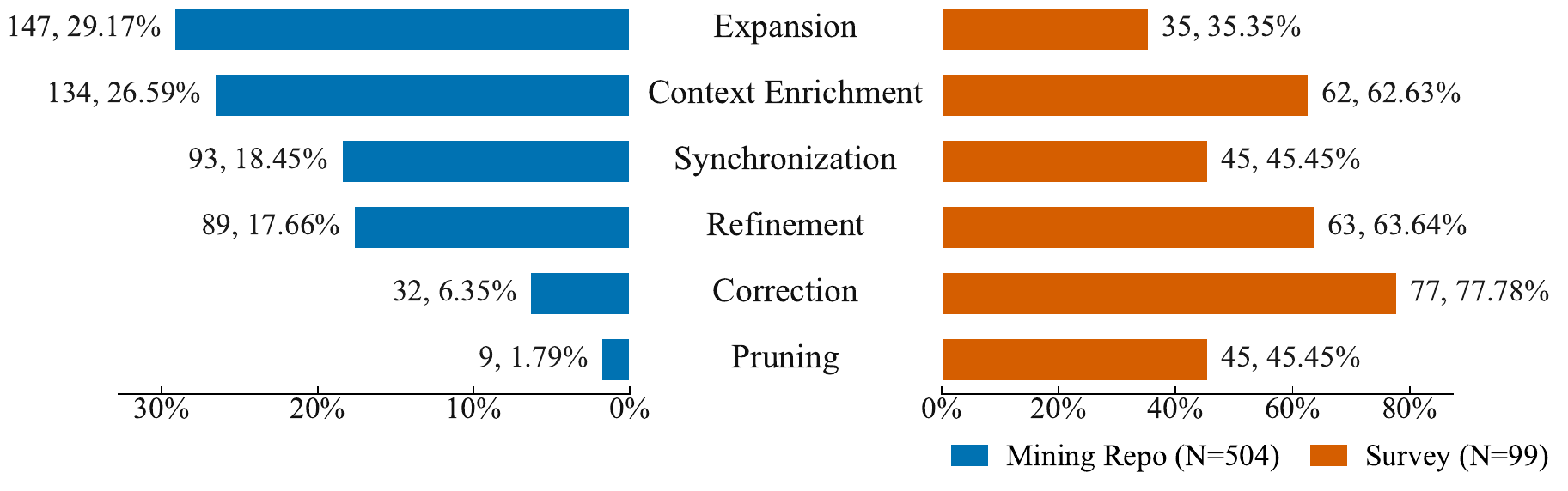}
\caption{Triangulation of driving reasons for rule evolution: mining data vs. survey responses.}
\Description{A butterfly chart comparing the distribution of rule evolution reasons. The left panel shows objective mining data (N=504), skewing heavily towards ``Expansion'' (29.17\%) and ``Context Enrichment'' (26.59\%). The right panel shows subjective survey responses (N=99), skewing dramatically in the opposite direction towards ``Correction'' (77.78\%) and ``Refinement'' (63.64\%).}
\label{fig:rq2_2_reason_mining_survey}
\end{figure*}

\textbf{Triangulation of Driving Reasons between Repositories and Surveys.}
To examine whether developers' perceived reasons correspond with the mining data, we compared the distribution of the mined data with the survey responses (Q10, multiple-choice) using a butterfly chart (Figure \ref{fig:rq2_2_reason_mining_survey}). 
The data reveals a clear difference between mined reasons and surveyed triggers. In the mining data (left), rule evolution is mostly driven by constructive purposes, where \textit{Expansion} (29.17\%) and \textit{Context Enrichment} (26.59\%) account for more than half of the changes. In contrast, \textit{Correction}-driven changes are less frequent in the mining data (6.35\%). However, the survey data (right) presents a different pattern: respondents frequently select \textit{Correction} (77.78\%) and \textit{Refinement} (63.64\%) as their primary reasons for modifying rules.

\textbf{Mapping Reasons to Change Types.}
We further investigated the relationship between the reasons for rule evolution and their corresponding change types (Table \ref{tab:rq2_2_reason_vs_changetype}). The data shows a clear association between specific reasons and change types. 
\textit{Added} is common among constructive reasons, accounting for 82.99\% of \textit{Expansion} and 84.33\% of \textit{Context Enrichment}. Additionally, \textit{Correction} is mostly performed through addition (68.75\%), suggesting that developers often fix AI hallucinations by adding new rules specifying negative constraints rather than modifying existing ones. \textit{Modified} dominates in \textit{Synchronization} (60.22\%) and \textit{Refinement} (57.30\%). As expected, all \textit{Pruning} cases (100\%) involve \textit{Deleted}.

\begin{table}[htbp]
    \centering
    \caption{Cross-Tabulation of Rule Evolution Reasons vs. Change Types (Mining Repo, $N=504$)}
    \label{tab:rq2_2_reason_vs_changetype}
    \resizebox{0.99\linewidth}{!}{
    \begin{tabular}{l@{\quad}c@{\qquad}rr@{\qquad}rr@{\qquad}rr}
        \toprule
        \multirow{2}{*}{\textbf{Driving Reason}} & \multirow{2}{*}{\textbf{Total ($N$)}} & \multicolumn{2}{c}{\textbf{Added}} & \multicolumn{2}{c}{\textbf{Modified}} & \multicolumn{2}{c}{\textbf{Deleted}} \\
        \cmidrule(r{2em}){3-4} \cmidrule(r{2em}){5-6} \cmidrule{7-8}
         & & \# & \% & \# & \% & \# & \% \\
        \midrule
        Expansion          & 147 & 122 & 82.99\% & 25 & 17.01\% & 0 & 0.00\% \\
        Context Enrichment & 134 & 113 & 84.33\% & 21 & 15.67\% & 0 & 0.00\% \\
        Synchronization    & 93  & 29  & 31.18\% & 56 & 60.22\% & 8 & 8.60\% \\
        Refinement         & 89  & 32  & 35.96\% & 51 & 57.30\% & 6 & 6.74\% \\
        Correction         & 32  & 22  & 68.75\% & 7  & 21.88\% & 3 & 9.38\% \\
        Pruning            & 9   & 0   & 0.00\%  & 0  & 0.00\%  & 9 & 100.00\%\\
        \bottomrule
    \end{tabular}
    }
\end{table}

\begin{figure*}[t]
\centering
\includegraphics[
    width=0.8\textwidth,
]{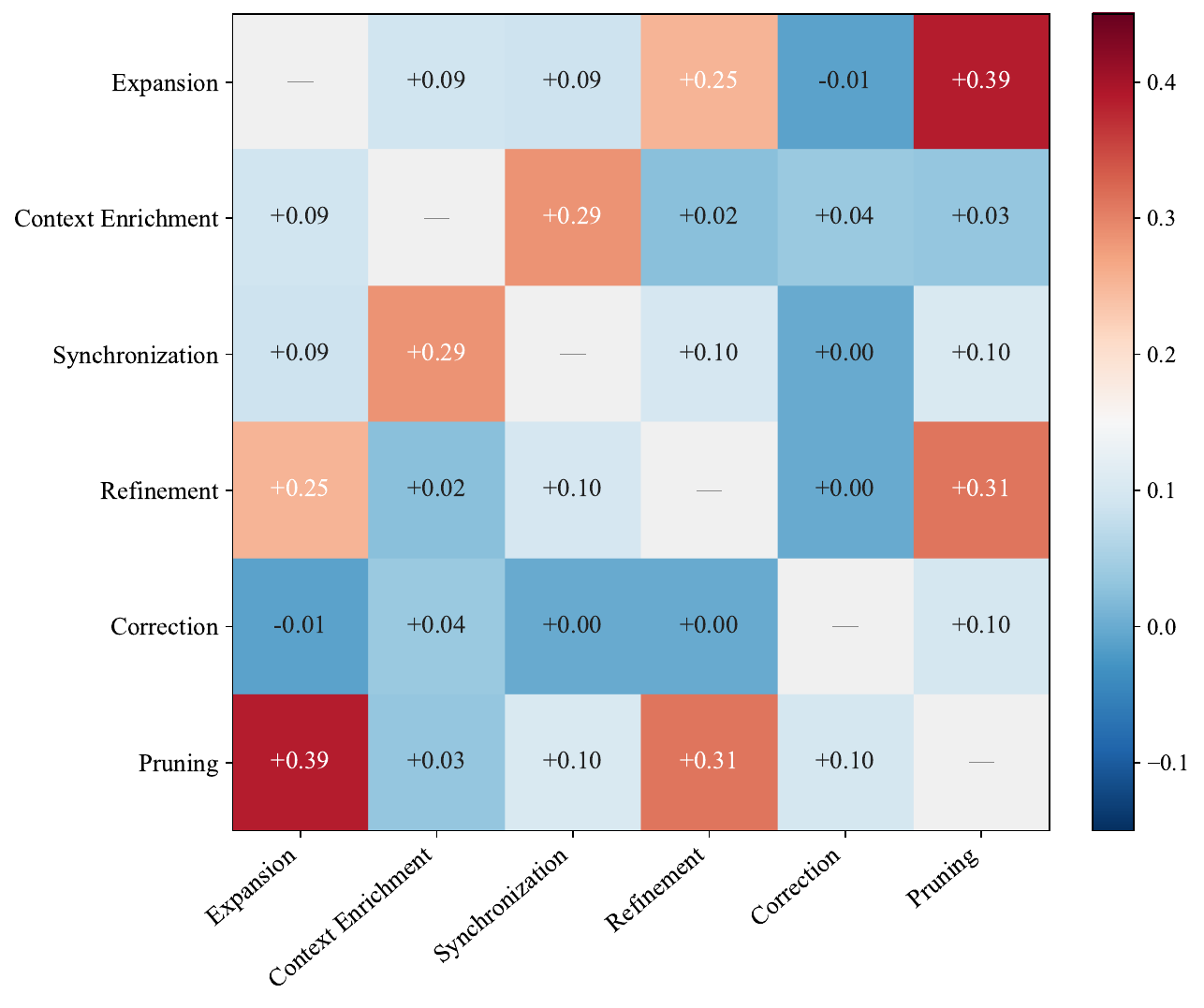}
\caption{Co-occurrence heatmap of subjective driving reasons based on Phi coefficients.}
\Description{A heatmap displaying the Phi correlation coefficients between 6 subjective reasons. The colors range from blue (negative) to dark red (positive). Three pairs show distinct positive correlations (darker red): Expansion and Pruning (+0.39), Refinement and Pruning (+0.31), and Synchronization and Context Enrichment (+0.29).}
\label{fig:rq2_2_survey_association_heatmap_phi}
\end{figure*}

\textbf{Co-occurrence and Statistical Testing of Surveyed Reasons.}
Finally, using  the multiple-choice survey data, we analyzed the co-occurrence patterns of these six reasons (Figure \ref{fig:rq2_2_survey_association_heatmap_phi}). Figure \ref{fig:rq2_2_survey_association_heatmap_phi} presents a heatmap of Phi ($\phi$) coefficients \cite{Pr2012}, where each cell shows the strength of association between any two reasons. A positive value indicates that the two reasons are selected together more frequently than expected by chance, while a value near zero implies they are independent. Treating each reason as a binary variable, we constructed 2$\times$2 contingency tables and used Fisher's exact test (two-sided) to evaluate independence \cite{Ki2017}. To control the false discovery rate (FDR) across the 15 pairwise comparisons \cite{St2017}, we applied the Benjamini-Hochberg (BH) procedure ($\alpha = 0.05$) \cite{ChSa2020}. After FDR correction, three pairs show statistically significant positive associations ($q_{\mathrm{BH}} < 0.05$):
\begin{itemize}
    \item \textit{Expansion} and \textit{Pruning} ($\phi = 0.39, q_{\mathrm{BH}} = 0.002$)
    \item \textit{Refinement} and \textit{Pruning} ($\phi = 0.31, q_{\mathrm{BH}} = 0.023$)
    \item \textit{Synchronization} and \textit{Context Enrichment} ($\phi = 0.29, q_{\mathrm{BH}} = 0.032$)
\end{itemize}

The $\phi$ values (ranging from approximately 0.29 to 0.39) indicate moderate positive correlations. Although \textit{Correction} has a high selection frequency, its co-occurrence with other reasons does not pass the significance threshold, suggesting that a high selection frequency does not necessarily imply a stronger statistical association within developers' perceptions.

\begin{tcolorbox}[
    enhanced,
    colback=gray!12!white,
    colframe=gray!50!black,
    arc=1.5mm,
    boxrule=1.4pt, 
    left=4pt, right=4pt, top=4pt, bottom=4pt
]
    \textbf{Key Finding 3:} There is a notable difference between the reasons for rule evolution and developers' perceptions: mining repository data shows that rule changes are mostly driven by \textit{Expansion} and \textit{Context Enrichment}, whereas survey respondents highlight \textit{Correction} and \textit{Refinement} as their primary triggers. Furthermore, while reasons like \textit{Expansion} and \textit{Pruning} frequently occur simultaneously, correcting AI errors typically happens in isolation by simply adding new rules.
\end{tcolorbox}

\subsubsection{RQ2.3: How does the compliance rate of software artifacts change following rule evolution?}\label{sec_rq2.3_result}
To answer this question, we compared the mining data with developer feedback from the survey. First, to investigate the temporal trends between rule evolution and software artifacts, we conducted a longitudinal analysis tracking compliance rates across 5 commits before and after a rule change. Subsequently, we supplemented these measurements with survey results regarding the practical impacts of rule maintenance.

\begin{figure*}[t]
\centering
\includegraphics[
    width=\textwidth,
]{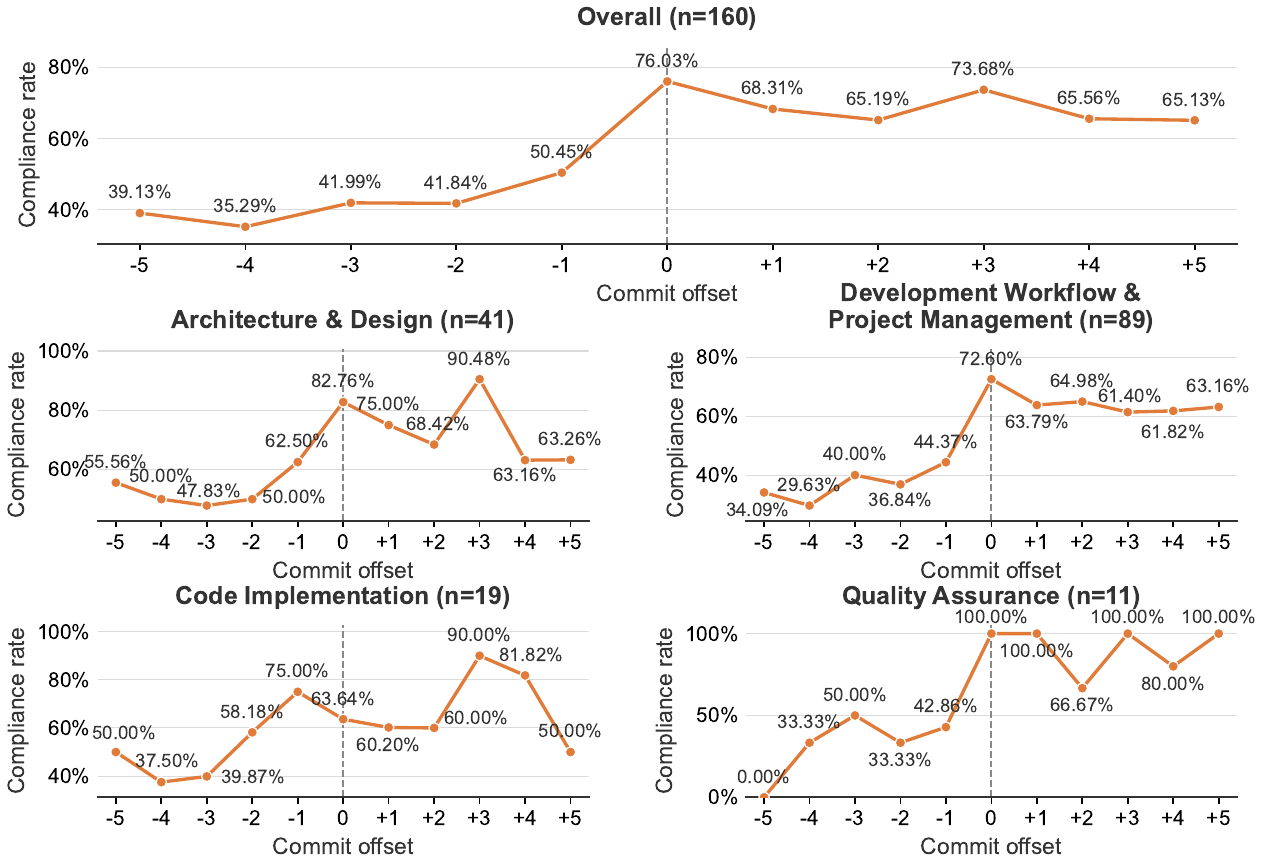}
\caption{Longitudinal trends of artifact compliance rates across 5 commits before and after rule evolution.}
\Description{A grid of five line charts showing compliance rates from commit offset -5 to +5. The 'Overall' chart shows compliance hovering between 35\% and 50\% prior to evolution, dramatically leaping to 76.03\% at commit 0, and stabilizing above 65\% in subsequent commits. Similar immediate leaps at commit 0 are observed in the sub-charts for the primary categories: Architecture \& Design, Development Workflow \& Project Management, Code Implementation, and Quality Assurance.}
\label{fig:rq2_3_first_level_compliance}
\end{figure*}

\textbf{Longitudinal Trends over Commits.}
To evaluate the effect of rule evolution, Figure \ref{fig:rq2_3_first_level_compliance} illustrates the trend of compliance rates for 160 rules across 5 commits before and after the change event ($Commit_0$). The \textit{Overall} trend shows an intervention effect: compliance fluctuates between 39.13\% and 50.45\% prior to the evolution (offsets -5 to -1), increases to 76.03\% at the commit of the change ($Commit_0$), and remains above 65\% in the subsequent +1 to +5 offsets. Across these 160 rules, the average compliance rate reaches 72.13\% after evolution, representing an increase of 22.99\% compared to the pre-evolution average of 49.14\%. This improvement is statistically significant based on a Wilcoxon signed-rank test ($p < 0.001$, $r = 0.71$) \cite{Wo2007}. This increase at $Commit_0$ occurs across the primary rule categories. For instance, compliance in \textit{Architecture \& Design} increases from 62.50\% at $Commit_{-1}$ to 82.76\% at $Commit_0$, and reaches 90.48\% at $Commit_{+3}$. Similarly, \textit{Development Workflow \& Project Management} increases from 44.37\% at $Commit_{-1}$ to 72.60\% at $Commit_0$. Although \textit{Code Implementation} shows fluctuations after the update, it reaches a peak of 90.00\% at $Commit_{+3}$. Finally, \textit{Quality Assurance} reaches 100.00\% at $Commit_0$ and maintains this rate in subsequent commits.

\begin{figure*}[t]
\centering
\includegraphics[
    width=\textwidth,
]{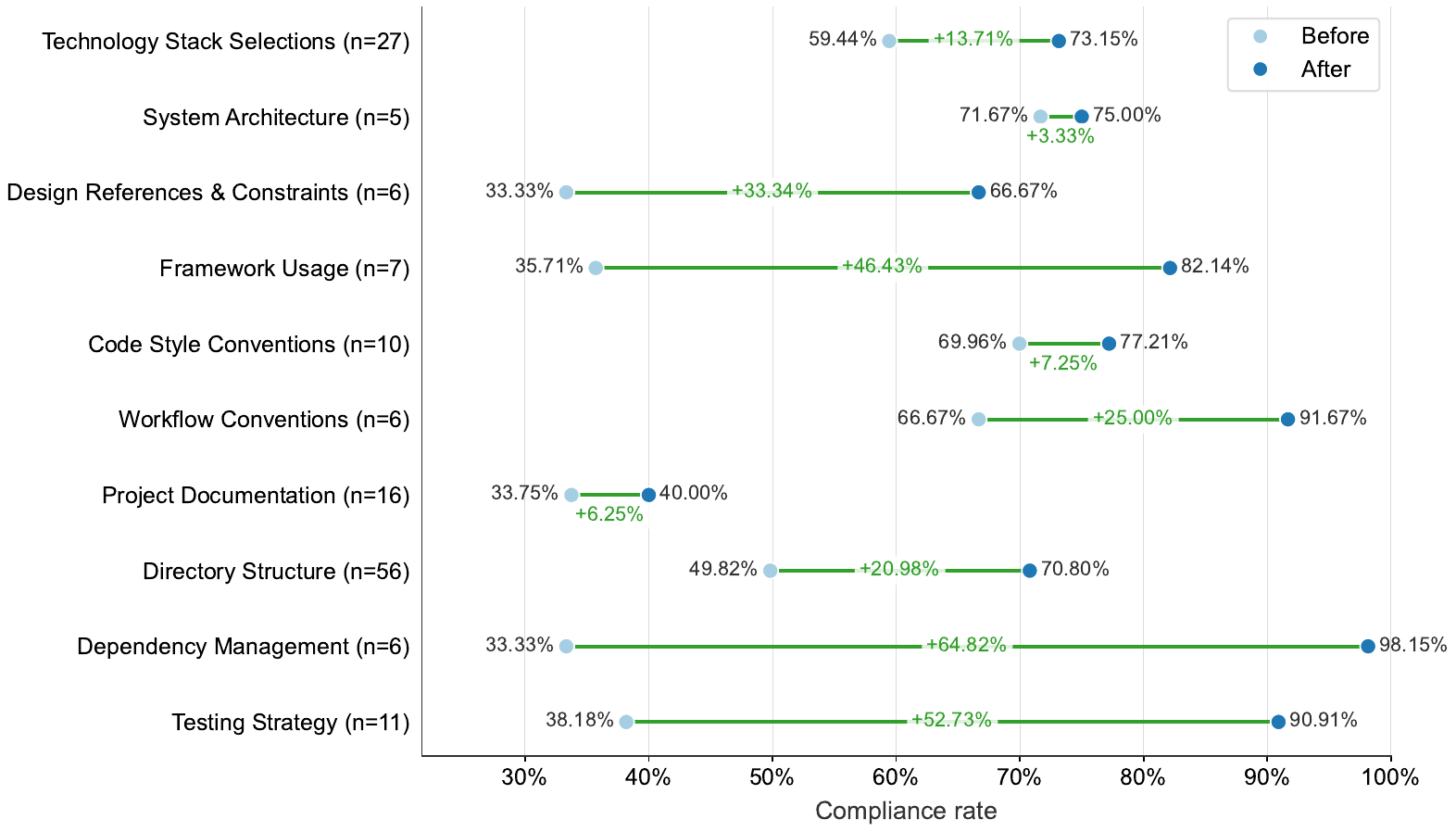}
\caption{Changes in average compliance rates for selected secondary categories before and after rule evolution ($n \ge 5$).}
\Description{A dumbbell plot contrasting pre-evolution and post-evolution compliance rates across 10 secondary categories. 'Dependency Management' displays the largest absolute increase (+64.82\%), jumping from 33.33\% to 98.15\%. 'Testing Strategy' and 'Framework Usage' also show massive gains exceeding 45\%. In contrast, 'System Architecture' and 'Code Style Conventions' exhibit minor increases below 8\%, and 'Project Documentation' shows a minimal +6.25\% increase from a severe low baseline of 33.75\%.}
\label{fig:rq2_3_second_level_compliance}
\end{figure*}

\textbf{Variations in Category-Specific Changes.}
To analyze these changes at a more granular level, Figure \ref{fig:rq2_3_second_level_compliance} contrasts the compliance changes across selected secondary rule categories. The data shows differences in how various types of rules affect compliance of software artifacts. Certain categories show immediate improvements; for example, \textit{Dependency Management} increases by 64.82\%, from 33.33\% to 98.15\%. This is followed by \textit{Testing Strategy} and \textit{Framework Usage}, which increase by 52.73\% and 46.43\%, respectively. Conversely, other categories show smaller changes. For instance, \textit{Code Style Conventions} and \textit{System Architecture} increase by 7.25\% and 3.33\%, respectively, as their pre-evolution baselines are already relatively high at 69.96\% and 71.67\%. Notably, \textit{Project Documentation} starts with a baseline of 33.75\% and shows a small increase of 6.25\% after the update, reaching 40.00\%. This suggests that modifying rule files is less effective for guiding AI adherence to macro-level documentation standards.

\begin{table*}[htbp]
    \centering
    \caption{Practical Impacts of Maintaining and Updating AI IDE Rules ($N=99$, Multiple Choice)}
    \label{tab:rule_impacts}
    \footnotesize 
    \begin{tabularx}{\linewidth}{lXrr} 
        \toprule
        \textbf{Practical Impact} & \textbf{Description} & \textbf{\#} & \textbf{\%} \\
        \midrule
        Quality Improvement & {\footnotesize Significant improvements in architecture, logic correctness, or security (beyond just formatting).} & 79 & 79.80\% \\
        \addlinespace
        Code Style Consistency & {\footnotesize Ensures the generated code follows specific naming conventions and formatting standards, with limited impact on functional logic.} & 59 & 59.60\% \\
        \addlinespace
        Efficiency Only & {\footnotesize The code quality has not changed much, but the rules save developers from repeatedly typing context/prompts.} & 57 & 57.58\% \\
        \addlinespace
        Workflow Standardization & {\footnotesize Enforces project-specific processes (e.g., file placement, commit standards, and testing procedures).} & 50 & 50.51\% \\
        \addlinespace
        Output \& Behavioral Alignment & {\footnotesize Regulates AI's output format, tone, and decision logic to match user preferences.} & 44 & 44.44\% \\
        \addlinespace
        Ineffective / Negative & {\footnotesize The AI frequently ignores the rules, or the rules cause unintended side effects (e.g., logic errors or over-restriction).} & 18 & 18.18\% \\
        \addlinespace
        Other & {\footnotesize Custom usages. One respondent specifically utilizes rules as a diagnostic probe: embedding specific user titles to detect whether the AI's context window has compressed or lost critical information.} & 1 & 1.01\% \\
        \bottomrule
    \end{tabularx}
\end{table*}

\textbf{Practical Perceptions of Rule Evolution Impacts.} 
To investigate how developers perceive the impacts of maintaining these rules, we analyzed responses to a multiple-choice question (Q11) in our survey. Table \ref{tab:rule_impacts} summarizes these findings. The survey data indicates that \textit{Quality Improvement} is the most common benefit, selected by 79 respondents (79.80\%). This is consistent with the mining data showing that rule updates improve compliance. Developers also report benefits in \textit{Code Style Consistency} (59, 59.60\%) and \textit{Workflow Standardization} (50, 50.51\%). Additionally, 57 respondents (57.58\%) selected \textit{Efficiency Only}, noting that rule maintenance saves them from repeatedly entering context even without direct code quality improvements. Conversely, 18 respondents (18.18\%) reported \textit{Ineffective/Negative} impacts, noting cases where the AI IDE ignores rules or introduces unintended constraints. One respondent under the ``Other'' category used rules as a diagnostic tool, embedding specific titles to detect whether the context window omits critical instructions. Overall, the survey responses indicate that rule evolution provides several benefits, from architectural alignment to prompting efficiency, supporting the improvements in artifact compliance rates observed in the mining data.

\begin{tcolorbox}[
    enhanced,
    colback=gray!12!white,
    colframe=gray!50!black,
    arc=1.5mm,
    boxrule=1.4pt, 
    left=4pt, right=4pt, top=4pt, bottom=4pt
]
    \textbf{Key Finding 4:} Modifying rules leads to a statistically significant increase in the compliance of software artifacts, with the average compliance rate rising by 22.99\% after a rule update. However, this effect varies across categories: updates to \textit{Dependency Management} and \textit{Testing Strategy} rules effectively improve artifact compliance, while rules for \textit{Project Documentation} show little change in their compliance rates. These trends regarding artifact compliance in the mining data are consistent with our survey results, where most developers attribute improvements in code quality and efficiency to rule evolution.
\end{tcolorbox}

\section{Discussion}\label{sec_discussion}

\subsection{Interpretation of Study Results}\label{sec_discuss_interpretation}

\subsubsection{Difference Between Perceived Importance and Rule Prevalence} 
Our findings from RQ1 (Section~\ref{sec_rq1_result}) show a clear contrast between developers' importance ratings and the actual distribution of rules in the mining data (Table~\ref{tab:rule_taxonomy} and Figure~\ref{fig:scatter_rule_count_mean_score}). While high-level rules such as \textit{System Architecture} (4.25/5) and \textit{AI Context Management} (4.17/5) are rated as highly important by practitioners, they account for only 2.67\% and 1.76\% of the mined rules, respectively. In contrast, concrete rule categories such as \textit{Workflow Conventions} and \textit{Testing Strategy} are substantially more prevalent in the mined repositories, accounting for 9.29\% and 8.76\% of all rules, respectively. This difference can be explained by the common rule creation strategies (Table~\ref{tab:rule_creation_strategies}). Since the majority of developers rely on AI-assisted generation (71.72\%) or add rules incrementally when errors occur during daily tasks (48.48\%) \cite{BaJaPo2023}, the resulting rule files naturally reflect immediate, practical needs. Concrete development details like file structures, formatting preferences, and local workflow steps cause frequent, visible friction during code generation, prompting developers to add specific rules to mitigate this friction. In contrast, broader structural designs are more abstract and harder to translate into short, incremental instructions \cite{ScHeArCoFuKeLiKo2025}, leading to a lower number of architectural rules in the mining data despite their high perceived importance.

\subsubsection{Rule Modifications and the Deletion of Rules on Non-Functional Properties} 
The variations in rule change types (Figure~\ref{fig:rule_change_distribution}) reflect the inherent capability boundaries of current LLMs when handling tasks of different granularities. Structural rules like \textit{AI Output Content Guidelines} and \textit{Technology Stack Selections} exhibit high modification rates (37.50\% and 36.29\%, respectively), as these concrete constraints require frequent textual fine-tuning to guide AI generation or align with the evolving versions of third-party libraries and frameworks. Conversely, rules regarding \textit{Performance Optimization} and \textit{Security Practices} rarely change, yet their evolution is heavily dominated by deletions (65.22\% and 61.90\%, respectively), with \textit{Performance Optimization} showing a 0\% modification rate. This phenomenon happens because security and performance are inherently Non-Functional Requirements (NFRs), which are abstract and difficult to define precisely in natural language prompts. Attempting to enforce constraints that are both vague and strict can easily trigger over-defensive LLM behavior or hallucinated false positives \cite{Ni2026}, thereby impairing coding efficiency. As a result, developers often delete these abstract rules altogether and instead rely on traditional static analysis tools or dedicated testing frameworks to ensure artifact compliance \cite{PeAhTaDoKa2022}.

\subsubsection{Differences in Reasons for Rule Evolution: Mining Data vs. Survey Responses} 
We observe a clear difference between the reasons found in the mining data and the reasons reported by developers in the survey (Figure~\ref{fig:rq2_2_reason_mining_survey}). In the mining data, rule evolution is mostly driven by constructive purposes like \textit{Expansion} (29.17\%) and \textit{Context Enrichment} (26.59\%), while changes driven by \textit{Correction} are rare (6.35\%). However, the survey responses show an opposite trend, where developers frequently select \textit{Correction} (77.78\%) as their primary trigger. This gap highlights a classic cognitive negativity bias \cite{RoRo2001}: the friction and frustration of fixing AI generation errors leave a much stronger impression on developers than performing \textit{Expansion} or \textit{Context Enrichment} for new features. Furthermore, even when performing a \textit{Correction}, developers use \textit{Added} operations in 68.75\% of rules (Table~\ref{tab:rq2_2_reason_vs_changetype}). This result confirms that rule maintenance is characterized by continuous, additive patching; developers prefer to add new negative constraints to restrict AI behavior rather than refactor existing rule texts.


\subsubsection{Changes in Rule Compliance over Time} 
The results of RQ2.3 (Section~\ref{sec_rq2.3_result}) show that the compliance rate of software artifacts peaks immediately when a rule is introduced ($Commit_0$), reaching an average compliance rate of 76.03\% across the evaluated rules (Figure~\ref{fig:rq2_3_first_level_compliance}). Overall, the average compliance rate rises from 49.14\% to 72.13\% after the rule update, which is a statistically significant increase of 22.99\% ($p < 0.001, r = 0.71$). However, this compliance rate tends to decrease in subsequent commits, dropping to approximately 65\% at $Commit_{+4}$ and $Commit_{+5}$, although it remains higher than the pre-evolution compliance rates, which ranging from 39.13\% to 50.45\%. We suggest that this subsequent decline might be driven by rule staleness and the increasing length or complexity of the context window \cite{LiLiHePaBePeLi2024}. At $Commit_0$, developers focus their code modifications on the specific areas targeted by the rule, keeping the context clean. As development continues, the addition of business code and conversational logs increases the complexity of the context window, gradually weakening the model's attention \cite{Mc2026Instruction}. Furthermore, because the vast majority of rule changes are additions (50\% to 80\% in Figure~\ref{fig:rule_change_distribution}) rather than modifications to existing rules, static and outdated rules can eventually conflict with newly added code as the codebase evolves.

\subsection{Implications for Practitioners}\label{sec_implications_practitioners}
This section presents implications and recommendations for helping developers and team managers to author and maintain AI IDE rules more effectively.

\subsubsection{Shift Focus from High-level Rules to Specific Rules} 
Our findings from RQ1 show that although developers value high-level design constraints, actual rule files are dominated by low-level constraints, such as \textit{Code Style Conventions} (5.57\%) and \textit{Workflow Conventions} (9.29\%) (Table~\ref{tab:rule_taxonomy}). Additionally, the results of RQ2.1 indicates high modification rates for rules governing specific project setups, such as \textit{Technology Stack Selections} (36.29\%) and \textit{Directory Structure} (26.17\%) (Figure~\ref{fig:rule_change_distribution}). We recommend that developers delegate formatting and syntax checking to traditional front-end linters (e.g., Prettier, ESLint) instead of using natural language rule files in AI IDEs. Saving the limited context window for high-level constraints, such as domain-specific business logic or architectural patterns, allows AI IDEs to assist with structural design rather than basic code formatting \cite{LiLiHePaBePeLi2024}.

\subsubsection{Avoid Abstract Prompts for Non-Functional Requirements} 
The results of RQ2.1 show that rules for \textit{Performance Optimization} and \textit{Security Practices} have high deletion rates of 65.22\% and 61.90\%, respectively, with \textit{Performance Optimization} showing a 0\% modification rate (Figure~\ref{fig:rule_change_distribution}). This indicates that abstract descriptions of Non-Functional Requirements (NFRs) in AI IDE rules, such as ``write secure code'' and ``optimize performance'' are difficult for LLMs to follow through natural language prompts \cite{PeAhTaDoKa2022}. Developers should avoid vague instructions in rule files. Instead, they should either convert NFRs into concrete rules with explicit code patterns (e.g., ``always use parameterized queries'') or rely on static security testing tools to enforce these NFRs \cite{FiGh2026}.

\subsubsection{Regularly Refactor and Prune Rule Files} 
In the results of RQ2.2, we observed that developers often address AI generation errors by adding new rules (68.75\% of corrections use \textit{Added}) rather than modifying existing rules (Table~\ref{tab:rq2_2_reason_vs_changetype}). One potential risk of this approach is that it can cause rule files to become long and conflicting over time. However, the survey results show a positive association between \textit{Expansion} and \textit{Pruning} ($\phi = 0.39, q_{\mathrm{BH}} = 0.002$), which suggests that some developers remove old rules when introducing new ones (Figure~\ref{fig:rq2_2_survey_association_heatmap_phi}). We recommend that development teams regularly review their rule files to identify potential conflicts and inconsistencies. Rather than continuously adding new rules, developers should consolidate related instructions and remove obsolete constraints to keep rule configurations clear.


\subsubsection{Writing Specific Rules with Clear Conditions} 
The results for RQ2.3 indicate that general rules, such as \textit{Project Documentation}, exhibit a low artifact compliance increase of 6.25\% after a rule update (Figure~\ref{fig:rq2_3_second_level_compliance}). This indicates that broad rules do not effectively control LLMs' behavior \cite{PeHuSoCaRiAsGlMcIr2022}. Developers should write specific rules with clear trigger conditions instead of high-level statements. For example, rather than writing a general instruction like ``\textit{please follow the architectural specifications}'', a more effective rule would state ``\textit{every time files under \texttt{src/api} are modified, \texttt{docs/api.md} must be updated synchronously}''. When rules are associated with specific file paths or explicit code syntax, the AI IDE can follow the instructions more accurately.

\subsection{Implications for Researchers and AI IDE Builders}
\label{sec_implications_researchers}
This section identifies the limitations of current rule mechanisms in AI IDEs and outlines future research directions and tool iterations needed for both academia and industry.

\subsubsection{Automated Extraction of Rules} 
The findings of RQ1 show that while developers rate high-level rule categories like \textit{System Architecture} (4.25/5) and \textit{AI Context Management} (4.17/5) as important, they create few rules for them, with \textit{System Architecture} representing only 2.67\% of the collected rules (Table~\ref{tab:rule_taxonomy}). This contrast indicates that translating high-level design principles and context-handling strategies into precise natural language rules is more challenging than specifying explicit constraints such as code formatting or workflow procedures. Such knowledge (e.g., architectural knowledge~\cite{babar2009software}) is typically abstract, implicit, and distributed across the codebase, which limits its suitability for direct manual coding. This highlights an area for future work, where researchers can explore methods to automatically extract a project's architectural features and technical constraints from underlying software artifacts (e.g., dependency graphs, abstract syntax trees, and commit histories) \cite{ShLaTa2023}. By leveraging these extracted features, future AI IDEs can automatically generate natural language rule files. This automation enables repositories to incorporate the high-level architectural constraints that developers prioritize but rarely specify manually due to the effort required for extracting rules.

\subsubsection{Integrating Rules with Existing Static Analysis Tools} 
As observed in RQ1 and RQ2.1, developers frequently use context window space on formatting and style constraints, such as \textit{Code Style Conventions} (5.57\%) and \textit{Workflow Conventions} (9.29\%) (Table~\ref{tab:rule_taxonomy}). However, they often delete NFR-related rules like security (61.90\%) and performance (65.22\%) because LLMs do not reliably enforce these constraints through natural language prompts (Figure~\ref{fig:rule_change_distribution}). This indicates that future AI IDEs might not require developers to manually duplicate formatting guidelines in rule files. Instead, tools can integrate features that automatically parse existing configurations (e.g., \texttt{.eslintrc}, \texttt{tsconfig.json}, \texttt{pom.xml}) and convert them into rules constraints for LLMs. Furthermore, rather than relying on textual rules to specify non-functional properties, AI tools should be integrated with traditional front-end linters and static analysis tools to establish clear boundaries for LLM behavior.

\subsubsection{Support for Rule Refactoring and Conflict Detection} 
The findings in RQ2.2 show that developers mostly handle corrections by adding new rules (68.75\% of corrections use \textit{Added} operations) rather than editing existing ones (Table~\ref{tab:rq2_2_reason_vs_changetype}). This incremental addition can cause rule files to become long and introduce contradictions over time. To address this, AI IDE builders need to develop automated consistency-checking and deduplication mechanisms tailored specifically for rule files. Future AI IDEs might also provide rule refactoring features that can detect semantic conflicts, such as contradictory library instructions. In addition, by examining AI interaction logs, these tools could identify unused or outdated rules and help developers better manage their configurations.

\subsubsection{Decomposing High-Level Developer Intent} 
By comparing the survey results with the mining data, we identified a gap between developer expectations and actual tool performance (Figures~\ref{fig:rq2_2_reason_mining_survey} and~\ref{fig:rq2_3_second_level_compliance}). While developers place high value on architecture and design rules (scoring 4.16 out of 5 in the survey), our compliance analysis in RQ2.3 shows that AI IDEs achieve a low improvement in artifact compliance when enforcing these abstract rules, such as a 3.33\% increase in the compliance rate with \textit{System Architecture} rules (Figure~\ref{fig:rq2_3_second_level_compliance}). This gap suggests that development environments should move beyond inserting static rule text into the context window. Instead, AI IDEs need to integrate intent decomposition features that can automatically capture a developer's high-level goals and translate them into low-level, atomic rule constraints that LLMs can follow more accurately.

\section{Threats to Validity}\label{sec_validity}

\subsection{Internal Validity}\label{sec_internal_validity}
Internal validity concerns the rigor of our analysis and the potential for confounding factors in our study. 

\textit{Qualitative coding bias.} The open coding process used to construct the rule taxonomy (RQ1) and categorize evolution reasons (RQ2.2) involves subjective judgment. To mitigate this, we employed a negotiated agreement approach \cite{CaQuOsPe2013}, where two authors independently coded the data and resolved discrepancies through continuous discussion. The Cohen's Kappa scores achieved during our pilot phases indicate a strong consensus between the coders. 

\textit{Reliability of LLM-based filtering.} Relying on LLMs as judges for data filtering introduces risks related to hallucinations and reasoning errors. To minimize this risk, we refrained from relying on a single model; instead, we employed an ensemble of three models (Gemini 3 Flash Preview, GLM 5, and Qwen3 Max) and applied a majority voting mechanism. Furthermore, we designed conservative prompts and validated the LLM outputs against human-labeled ground truth to ensure evaluation accuracy.

\textit{Survey participant selection.} Randomly distributing questionnaires could introduce demographic bias. To address this, we mined the email addresses of Git commit authors who modified AI IDE rule files in OSS repositories. This step ensures that the respondents possess practical experience with rule configuration.

\textit{Causality in compliance assessment.} The longitudinal analysis in RQ2.3 aims to observe the temporal trends of rule evolution rather than establish causality between rule updates and subsequent changes in artifact compliance. In practice, an improvement in the compliance of software artifacts might result from the AI IDE adhering to the updated rules, but it could also stem from manual code adjustments made by developers within or around the same commit.

\textit{Selection bias in evaluated rules.} To ensure the reliability of the automated compliance assessment, we implemented filtering criteria. Specifically, we only retained rules whose artifact compliance could be objectively and accurately measured via automated scripts, prioritizing evaluation precision over sample recall. This filtering process reduced the final evaluated sample to 160 rules, introducing selection bias. The remaining rules mostly belong to statically verifiable categories (e.g., \textit{Directory Structure} and \textit{Dependency Management}). Consequently, the observed compliance improvements reflect how well AI IDEs follow concrete structural constraints specified in rules, and may not generalize to more abstract rules.

\subsection{Construct Validity}\label{sec_construct_validity}
Construct validity assesses whether our experimental design and metrics accurately capture the constructs we intend to study.

\textit{Repository identification.} Our repository collection relied on specific keywords (Table \ref{tab:keywords}) and file extensions (Table \ref{tab:ai_ides}). This approach misses projects that use AI IDEs but do not explicitly declare them in their README files. Given GitHub API limitations, this is a common compromise in mining repositories studies \cite{DaAgBa2021}. The 83 manually verified projects represent a sample of current AI IDE usage.

\textit{Rule segmentation.} Segmenting natural language Markdown files into individual rules introduces boundary ambiguity. To standardize this process, we established criteria for defining a ``semantic unit''. The first author conducted the segmentation, and the co-authors audited the results to ensure consistent granularity.

\textit{Survey response bias.} Asking survey participants to rate the importance of 25 rule categories on a Likert scale may trigger a ``ceiling effect'', where respondents tend to assign high ratings across all items. To mitigate this bias and improve understandability, we conducted two rounds of pilot surveys to refine the phrasing. In our data analysis, we supplemented the absolute scores with Spearman's rank correlation (see Section \ref{sec:rq1_quantitative_analysis}) and quartile-based scatter analysis (Figure \ref{fig:scatter_rule_count_mean_score}) to draw relative conclusions.

\textit{Authorship of the analyzed code.} A core assumption of our study is that the code within the analyzed repositories is generated or heavily influenced by AI IDEs. Verifying the exact authorship (human vs. AI) of every line of code is not feasible. To mitigate this threat, we included a manual inspection step during data collection. We retained projects only if the developers explicitly declared in the README or repository description that the project was ``fully or primarily developed by AI''. This inclusion criterion ensures that our dataset reflects AI-assisted development practices.

\textit{Branch selection for evolution analysis.} To construct a consistent evolution history, our analysis was limited to the \texttt{main} or \texttt{master} branches of the repositories. While this approach captures the long-term, merged rule evolutions, it misses temporary rule modifications that occur within pull requests or feature branches. The evolution patterns we report reflect the stabilized maintenance of rules rather than intermediate iterations during drafting.

\subsection{External Validity}\label{sec_external_validity}
External validity refers to the generalizability of our findings.

\textit{Dataset representativeness.} Our collected dataset exhibits a representativeness bias: it primarily consists of TypeScript and Web development projects (Figure \ref{fig:language_and_domain}) created by solo developers or small teams. As a result, our findings may not generalize to large-scale enterprise systems (e.g., legacy C++ or Java backends). Nevertheless, these demographics reflect the current landscape of early adopters of AI-enabled IDEs.

\textit{Tool and ecosystem evolution.} The ecosystem of AI-assisted software engineering is evolving rapidly. Both the underlying foundation models and the rule-injection mechanisms of AI IDEs undergo frequent updates. Consequently, specific phenomenon observed during our data collection window may change over time. Despite this, our study provides a snapshot of current ``Context Engineering'' practices in AI-assisted development \cite{SeMaChSe2026}. The challenges we identified in how developers constrain AI behaviors through natural language rules provide a foundation for future research on rule-based harness engineering~\cite{harness} for agentic programming.

\subsection{Reliability}\label{sec_reliability_validity}
Reliability concerns the extent to which other researchers can independently replicate our study. 

\textit{LLM non-determinism and study replicability.} To address the non-determinism of LLM outputs, we configured the LLM API calls with structured prompts and enforced JSON output schemas. To support replication, we made our research artifacts publicly available in an online replication package \cite{dataset}. This repository includes the filtered datasets, qualitative codebooks, prompt templates, LLM responses, and statistical scripts, enabling external researchers to reproduce our data processing and analytical pipeline.

\section{Conclusions and Future Work}\label{sec_conclusion}
As a configuration mechanism in agentic programming, AI IDE rules guide AI behavior in modern software development. In this paper, we presented a mixed-methods empirical study on AI IDE rules. By mining 7,310 rules from 83 open-source repositories developed by AI IDEs and triangulating the data with survey feedback from 99 practitioners, we established a taxonomy comprising 5 primary and 25 secondary categories. Our findings reveal a contrast between developer priorities and actual rule configurations: while developers highly value architectural guidelines, repositories primarily contain low-level formatting and workflow rule constraints. Regarding rule evolution, we observed that rules are regularly updated during development. While repository data shows that evolution is often driven by constructive context expansions, developers report modifying rules mainly to correct AI generation errors, typically by adding new rules rather than refactoring existing ones. Finally, our artifact compliance analysis shows that updating rules improves the compliance of software artifacts. However, this improvement is more pronounced for concrete, statically verifiable constraints than for abstract guidelines in rules. These findings shed light on current rule management practices in AI IDEs and provide an empirical foundation for optimizing rule-based configurations and maintenance strategies.

Based on the limitations identified in our study, future research could be extended in the following three directions: 
(1) \textit{Conflict detection and rule management.} Since developers often accumulate negative constraints by adding new rules, future work could develop maintenance tools tailored to natural language prompts. Such tools could detect semantic conflicts and identify unused rules, helping developers maintain concise rule configurations.
(2) \textit{Automated extraction of architectural rules.} To reduce the manual effort required to author high-level design rules, researchers could explore techniques for automatically reverse-engineering and generating high-level constraints from sources such as codebase abstract syntax trees, dependency graphs, and commit histories.
(3) \textit{Large-scale enterprise validation.} The sample collected in this study primarily consists of Web development and small-to-medium-sized projects, reflecting the current early-adopter ecosystem of AI IDEs. Future studies could extend our methodology to investigate rule maintenance patterns and longitudinal artifact compliance trends in more complex, large-scale enterprise environments, such as legacy C++ or Java backend systems.

\section*{Data Availability} The replication package of this study has been made available at \cite{dataset}.

\section*{Acknowledgments}
This work has been partially supported by the National Natural Science Foundation of China (NSFC) with Grant No. 92582203 and 62402348.

\bibliographystyle{ACM-Reference-Format}
\bibliography{reference}

\balance

\end{document}